\documentclass[5p]{elsarticle}

\usepackage{amssymb}
\usepackage{epsfig}
\usepackage{graphics}
\usepackage{graphicx}
\usepackage{subfigure}
\usepackage{exscale}
\usepackage{latexsym}
\usepackage{makeidx}
\usepackage[section]{placeins}
\usepackage{xtab}
\usepackage{epsf}
\usepackage{bbm}
\usepackage{color}

\usepackage{lineno}

\journal{Nuclear Instruments and Methods in Physics Research Section A}

\begin{document}
\begin{frontmatter}

\title{Measurement of neutron yield by 62 MeV proton beam on a thick Beryllium target}

\author{M.~Osipenko$^a$, M.~Ripani$^a$, R.~Alba$^b$, G.~Ricco$^a$, M.~Schillaci$^b$,
M.~Barbagallo$^c$, P.~Boccaccio$^d$, A.~Celentano$^e$, N.~Colonna$^c$,
L.~Cosentino$^b$, A.~Del~Zoppo$^b$, A.~Di~Pietro$^b$, J.~Esposito$^d$, P.~Figuera$^b$,
P.~Finocchiaro$^b$, A.~Kostyukov$^f$, C.~Maiolino$^b$,
D.~Santonocito$^b$, V.~Scuderi$^b$, C.M.~Viberti$^e$
}

\address{
$^a$ \it\small INFN, sezione di Genova, 16146 Genova, Italy, \\
$^b$ \it\small INFN, Laboratori Nazionali del Sud, 95123 Catania, Italy, \\
$^c$ \it\small INFN, sezione di Bari, 70126  Bari, Italy, \\
$^d$ \it\small INFN, Laboratori Nazionali di Legnaro, 35020 Legnaro, Italy, \\
$^e$ \it\small Dipartimento di Fisica dell'Universit\`a di Genova, 16146 Genova, Italy, \\
$^f$ \it\small Moscow State University, Moscow 119992, Russia
}

\begin{abstract}
The design of a low-power prototype of neutron amplifier recently proposed within the INFN-E project
indicated the need for more accurate data on the neutron yield produced by a proton beam with energy of about 70 MeV 
impinging on a thick Beryllium target. Such measurement was performed at the LNS superconducting cyclotron, covering a wide angular range from 0 to 150 degrees and a complete neutron energy interval from thermal to beam energy.
Neutrons with energy above 0.5 MeV were measured by liquid scintillators exploiting their Time of Flight
to determine the kinetic energy. For lower energy neutrons, down to thermal energy, a $^3$He detector was used.
The obtained data are in good agreement with previous measurements
at 0 degree using 66 MeV proton beam, covering neutron energies $>$10 MeV,
as well as with measurements at few selected angles using protons of 46, 55 and 113 MeV energy.
The present results extend the neutron yield data in the 60-70 MeV beam energy range.
A comparison of measured yields to MCNP, FLUKA and Geant4 Monte Carlo simulations was performed.
\end{abstract}


\end{frontmatter}

\section{Introduction}\label{sec:intro}
Research on fast neutron reactors addresses, among others, the problems of the limited availability of
fissile materials and the difficulties in nuclear waste disposal. This is because fast neutrons induce fission
on larger number of nuclides, including $^{238}$U and Minor Actinides like e.g. $^{241}$Am,
while they undergo much less capture processes leading to the build up of
high toxicity waste. Accelerator Driven System (ADS)~\cite{ADS} represents a promising solution to this problem. It consists of a fast sub-critical reactor fed by an external neutron source of sufficiently high intensity, that can be generated by a particle accelerator or by a fusion reactor.

The high power proton cyclotron that will be installed at Laboratori Nazionali di Legnaro
of INFN for the SPES project~\cite{SPES} offers the possibility to build a low-power ADS prototype
for research purposes. The design of such prototype has been developed
by a broad collaboration between INFN and other bodies and is documented in Ref.~\cite{infn_e_ads}. It is based on a 0.5 mA, 70 MeV proton beam
impinging on a thick Beryllium target and a fast sub-critical core consisting of solid lead
matrix and 60 $UO_2$ fuel elements, enriched to 20\% with $^{235}$U. Both, production
target and reactor core, are cooled by continuous helium gas flow. The ADS is expected
to have the effective neutron multiplication factor $k_{eff}=0.946$,
neutron flux $\phi=3\div 6\times 10^{12}$ n/cm$^2$/s and thermal power of 130 kW at 200$^\circ$ C.
Such a facility would allow to study the kinetics and dynamics of the fast reactor core,
the burn-out and transmutation of radioactive waste, as well as issues related to system safety and licensing.

The choice of the production target material is determined by the low beam energy of
the cyclotron. $^9$Be bombarded by protons, in fact, provides an abundant neutron
source whose spectrum is sufficiently hard to burn minor actinides. The design of
the proposed ADS requires careful knowledge of the neutron yield produced by the 70 MeV
proton beam on a thick $^9$Be target. The existing data in the given energy range are rather
scarce and not comprehensive. The integrated yield at 70 MeV was measured in Ref.~\cite{Tilquin05},
while differential yields for few angles were measured at various beam energies
in Refs.~\cite{Waterman79,Johnsen76,Almos77,Heintz77,Meier88,Madey77,Harrison80}.
However, only the data from Ref.~\cite{Almos77} have sufficiently similar beam energy,
also these results are limited to 0 degrees and neutron energies $>$10 MeV.
This lack of data demanded a dedicated measurement of neutron yield produced by
a proton beam on a thick $^9$Be target. The measurement was performed
at the Laboratori Nazionali del Sud (LNS) of INFN using the existing superconducting cyclotron~\cite{cyclotron,lnsrep10}.
The cyclotron was usually operated with a 62 MeV proton beam, hence we also
selected this, well tuned, energy, close enough to the project goal.
Time of Flight in liquid scintillators was used to measure neutrons with energies $>$0.5 MeV,
while a $^3$He detector had been used for neutrons of lower energy.
The precision of the experiment was of the order of 10\%,
dominated by systematic uncertainties on the detector efficiencies and absolute normalization.

In the first three sections we discuss experimental setup, detector geometry and energy calibrations.
In further sections the data analysis is presented.
Finally in the last few sections the obtained data are discussed and compared
to previous data and Monte Carlo simulations.

\section{Experimental Setup}\label{sec:baf}
The 62 MeV proton beam extracted by the electrostatic deflector from the LNS superconducting cyclotron
was transported for 65 m through a series of quadrupole and 8 dipole magnets, rotating the beam by 270 degrees,
to be delivered in the experimental hall. All the beam transport system was optimized for 62 MeV protons.
The operating beam current was selected to be 30-50 pA.
The cyclotron beam structure (RF=40 MHz) was modified by suppression of four bunches out of five. This way 1.5 ns wide beam bunches arrived on the target
with period of 125 ns.
The target consisted of a solid 3 cm thick $^9$Be cylinder with 3.5 cm diameter. The target thickness
and width were chosen to ensure complete absorption of the proton beam.
The electric charge deposited by the beam on the target was measured by a digital current integrator
and used for absolute normalization of the data. Two different current integrators were used during the experiment for comparison.

\begin{figure}[h]
\begin{center}
\includegraphics[scale=0.35]{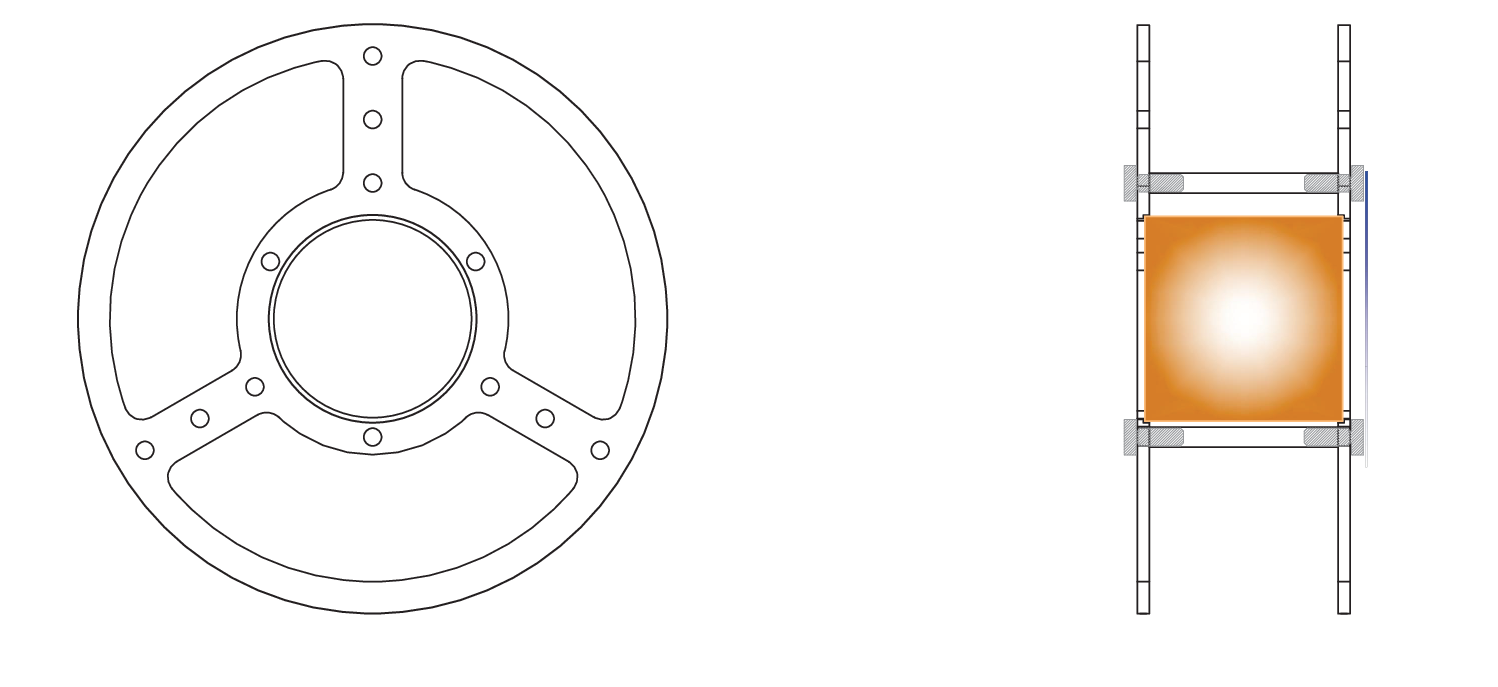}
\caption{\label{fig:target_holder}(color online) Target holder drawing. The beryllium target is represented by the yellow cylinder.}
\end{center}
\end{figure}

The precision of the present experiment depended on careful reduction of environmental background. To reduce $\gamma$ production and rescattering of the large flux of produced neutrons from surrounding materials specific measures were taken. First of all the beamline was located at a height of 168 cm above an empty floor. All beamline components downstream of the target were completely removed. Closest wall of the experimental hall was located at more than 4 m distance from the target. During preliminary measurements, the 3 mm thick steel beamline proved to be a significant source of background, in particular of n-$\gamma$ conversion. This was also confirmed by MCNP~\cite{mcnp5} simulations. To reduce this contamination a new, lightweight, final segment of the beamline was installed. It consisted of 100 cm long carbon fiber tube with 3 mm thick walls manufactured by Quick Batten~\cite{quickbatten}. The tube was connected to the main beamline via an ISO-K Aluminum flange glued to the carbon fiber. At the downstream side the tube was sealed by a glued 2 mm thick Aluminum endcap. Such solution permitted to reduce the background by more than a factor of 2, maintaining the vacuum quality better than $10^{-5}$ mbar.
The target holder was also specially designed to avoid large background production and good 
electric insulation to guarantee the best charge collection.
This was made of two Teflon wheels 2 mm thick with outer diameter equal to the beamline pipe size and inner diameter fitting the target cylinder as shown in Fig.~\ref{fig:target_holder}.
The front face of the target with respect to the beam had a guard ring polarized at -200 V
to prevent $\delta$-electron emission from the target. Tests made at different values of bias voltage
showed no significant variation of the collected charge.

\begin{figure}[h]
\begin{center}
\includegraphics[scale=0.28, angle=0]{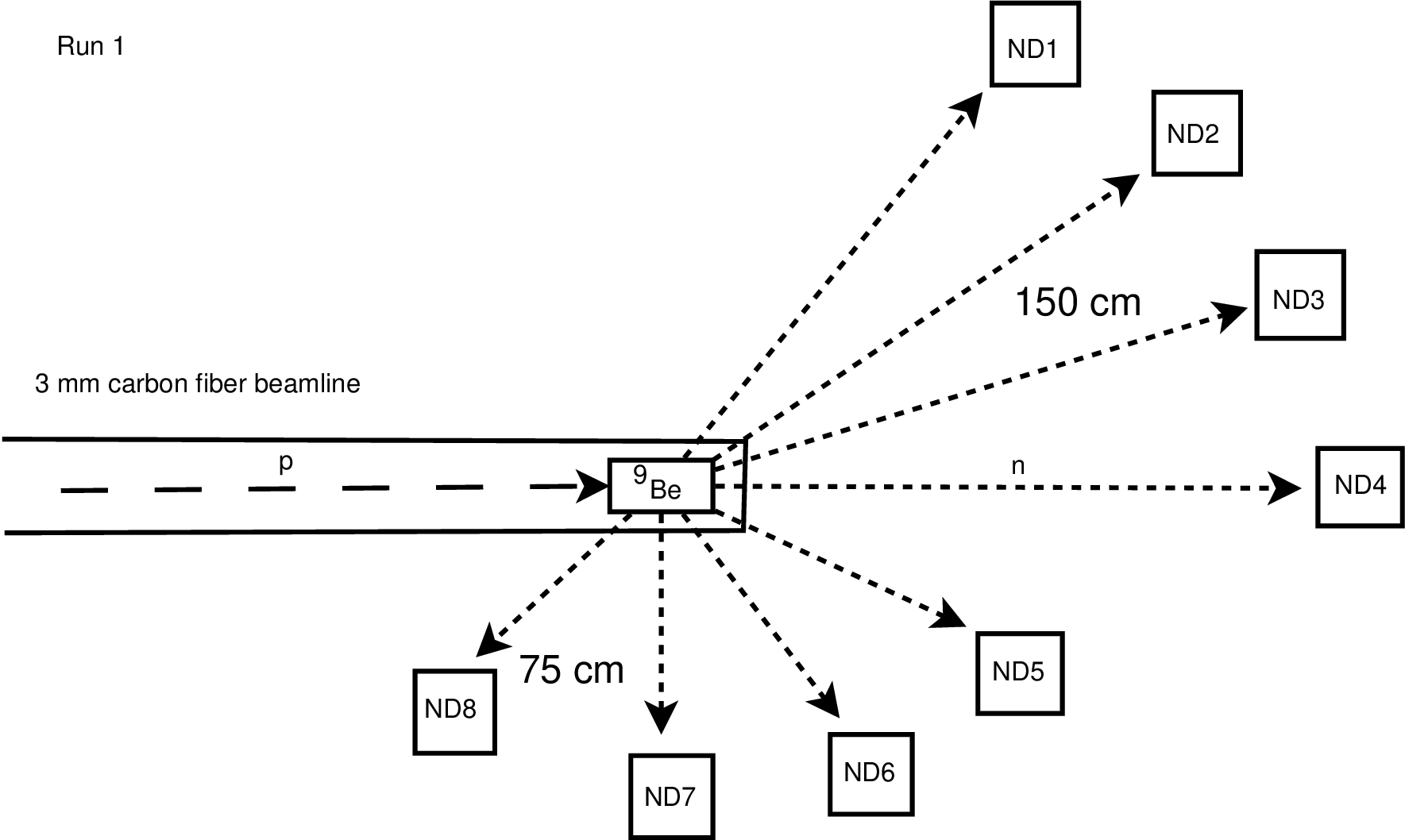}
\caption{\label{fig:setup} Schematic view of the experimental setup.
The detectors were moved simultaneously at other angles to cover
16 angular settings from 0 to 150 degrees.}
\end{center}
\end{figure}

The neutrons produced in the target were measured by the ToF technique. To this end, the RF signal from the cyclotron was used as the reference time. 
Typically, we obtained a time resolution of 1.6 ns (FWHM of the gamma peak).
Eight neutron detectors were installed simultaneously around the target, at the same height of the beamline at different angles and at two different distances (150 cm and 75 cm,
measured with 1 mm accuracy) around the target as shown in Fig.~\ref{fig:setup}. Many separate runs were taken during the experiment. For each run all eight detectors were moved simultaneously to other angles for a total of 16 positions, sometimes repeated for consistency checks.
Near (small) detectors, installed at half distance (75 cm) from the target with respect to far (large) detectors, were used to measure low energy neutrons ($T_n<$2 MeV).

All detectors were installed on light Aluminum tubes by using Teflon holders. The tubes were mounted on an Aluminum foot for mechanical stability. An additional support was foreseen in front of the detector to install a 50 cm long steel shadow bar designed for background measurements.

\section{Liquid scintillator detectors}

\subsection{Detector description}
\begin{figure}[!ht]
\begin{center}
\includegraphics[scale=0.80]{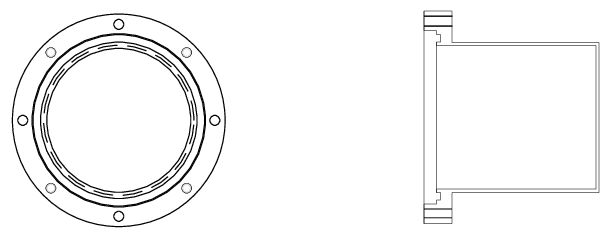}
\caption{\label{fig:cell_large}Drawing of large cell.}
\end{center}
\end{figure}

\begin{figure}[!ht]
\begin{center}
\includegraphics[scale=0.65]{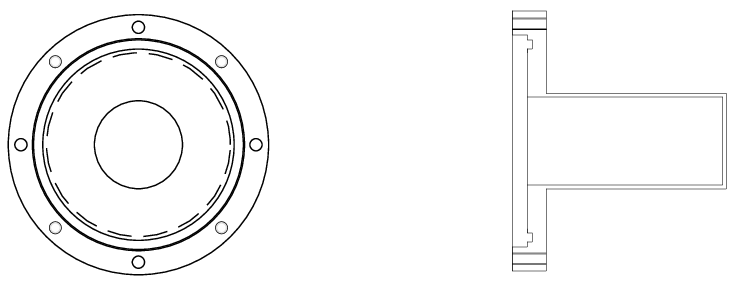}
\caption{\label{fig:cell_small}Drawing of small cell.}
\end{center}
\end{figure}

Neutron detectors were 4 cm long Aluminum cylindrical cells filled with liquid scintillator.
In order to measure neutron yields at two different distances from the target,
keeping the same solid angle, two different diameters were chosen.
Four cells used as far detectors had the diameter of 4.6 cm (large detectors)
while the other four, used as near detectors placed at half distance from the target,
had the diameter of 2.3 cm (small detectors).
The drawings of the two types of cells are shown in Figs.~\ref{fig:cell_large} and \ref{fig:cell_small}.


To enhance light collection, the cell Aluminum walls
were painted with white diffusive paint EJ520 (analog of NE-561 and BC-622A).
Small cells with Teflon diffuser were also produced for comparison.
Reflectivity of the paint varies from 83\% at 400 nm to 93\% at 520 nm.
Teflon was suggested to have somewhat better reflectivity~\cite{yamawaki:teflon},
however we did not observe any significant difference.

The cells were filled in Ar/N$_2$ atmosphere with EJ301 liquid scintillator,
having a good PSD capability.
The scintillator has density $\rho=0.874$ g/cm$^3$, atomic H:C ratio 1.212,
refractive index $n_D=1.505$, it produces 12,000 blue photons per 1 MeV electron equivalent (MeVee)
deposited energy with the first three excitation components decaying in 3.16, 32.3 and 270 ns.

The liquid scintillator did not fill the entire cell volume, leaving
a small bubble of Ar/N$_2$ for a possible thermal expansion.
The total amount of scintillator in each cell was measured
by a comparison of full to empty cell weights. The measured
expansion volume was compatible with the designed value of 4\%.

Each cell was sealed by a 4 mm thick borosilicate glass, coupled
to light sensor by optical grease.
As light sensors we selected Electron Tubes Enterprise PMTs 9954 with
high linearity voltage dividers C649. The choice was mainly determined
by a good overall quantum efficiency, extended in the red region. This was
particularly important for PSD~\cite{amaldi:psd_spectrum}.

\subsection{Calibrations}
In order to calibrate the signal of the liquid scintillators in terms of electron equivalent
deposited energy, the backward Compton scattering was used.
This reaction allows to observe a pronounced peak of Compton scattering
at 180 degrees and to remove underlying background by a fast coincidence. We used a BaF$_2$ crystal~\cite{BaF2} to detect the recoiled $\gamma$ at backward angles.
The distance between the BaF$_2$ crystal and the EJ301 cell was about 10-15 cm.

The measurements were performed at LNS with a number of $\gamma$ sources.
Small detectors covering the energy range up to 0.5 MeVee were calibrated with
$^{152}$Eu, $^{22}$Na, $^{207}$Bi and $^{137}$Cs sources.
For the large detectors, the energy scale was calibrated 
by means of $^{207}$Bi, $^{137}$Cs, $^{60}$Co and $^{22}$Na sources.

Events measured in coincidence between liquid scintillator
detector and BaF$_2$ crystal were analyzed off-line to select backward Compton scattering only.
The TDC distributions of the liquid scintillator-BaF$_2$ time-difference were analyzed in order to select coincidence events. Then we identified the corresponding peak of scattered $\gamma$ in BaF detector deposited energy and used the events in the peak for calibration. The charge distributions
of liquid scintillator signals were fitted by a Gaussian to determine the backward Compton scattering peak position and width.

For small detectors the 60 keVee photoelectric peak from an $^{241}$Am source was also
measured to cover the low energy domain. These measurements were found to be
in agreement with coincidence data.

\begin{figure}[!ht]
\begin{center}
\includegraphics[bb=2cm 6.5cm 20cm 23cm, scale=0.35]{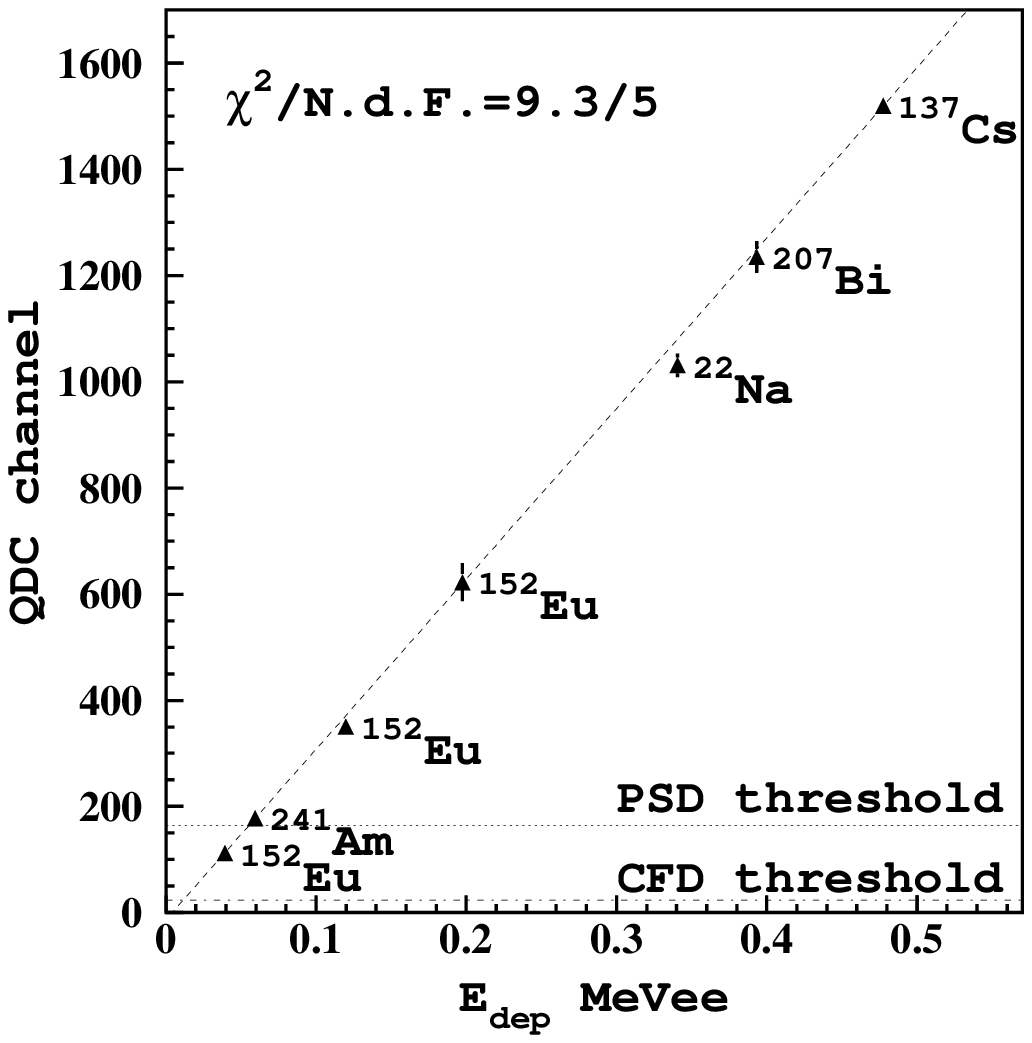}
\includegraphics[bb=2cm 6.5cm 20cm 24cm, scale=0.35]{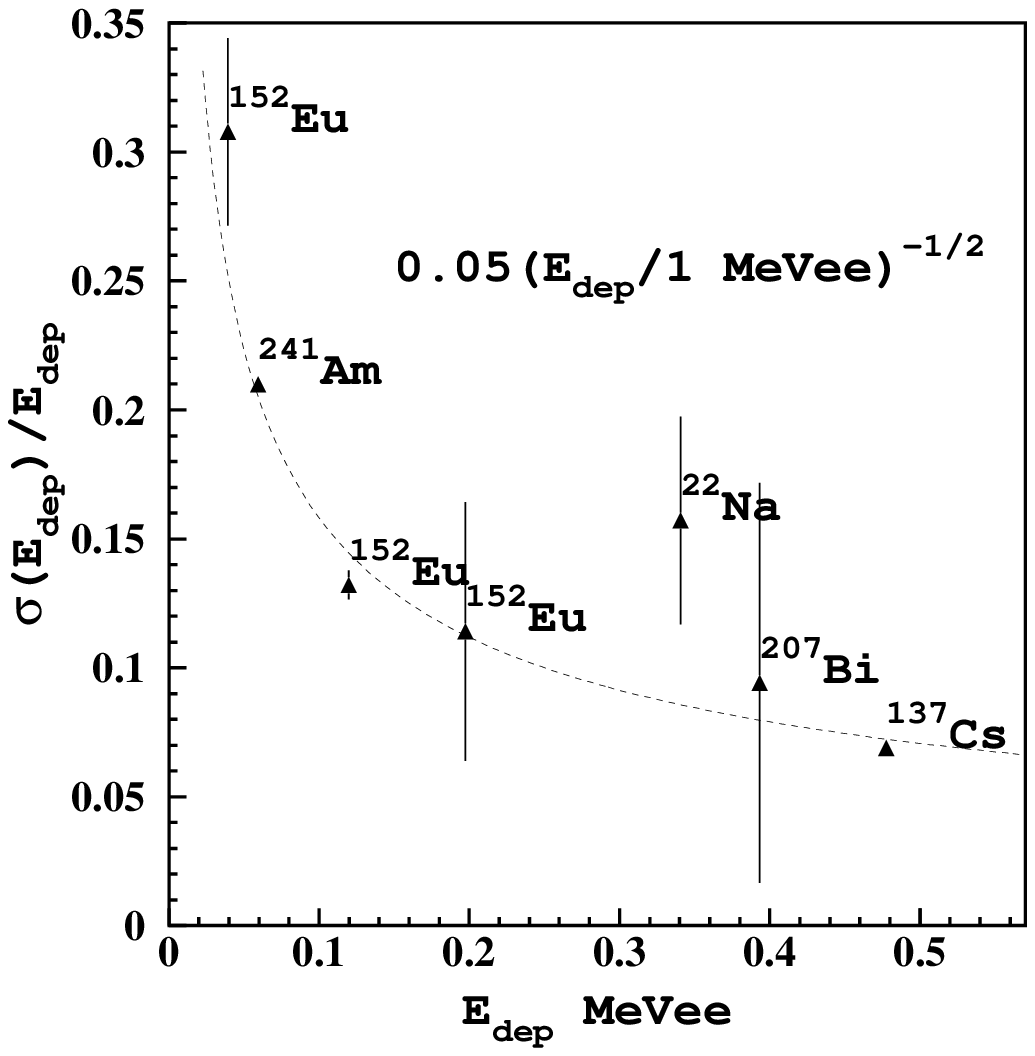}
\caption{\label{fig:nd6_qdc_calib}Small detector QDC calibration with $\gamma$ sources (upper)
and its energy resolution (lower).}
\end{center}
\end{figure}

\begin{figure}[!ht]
\begin{center}
\includegraphics[bb=2cm 6.5cm 20cm 23cm, scale=0.35]{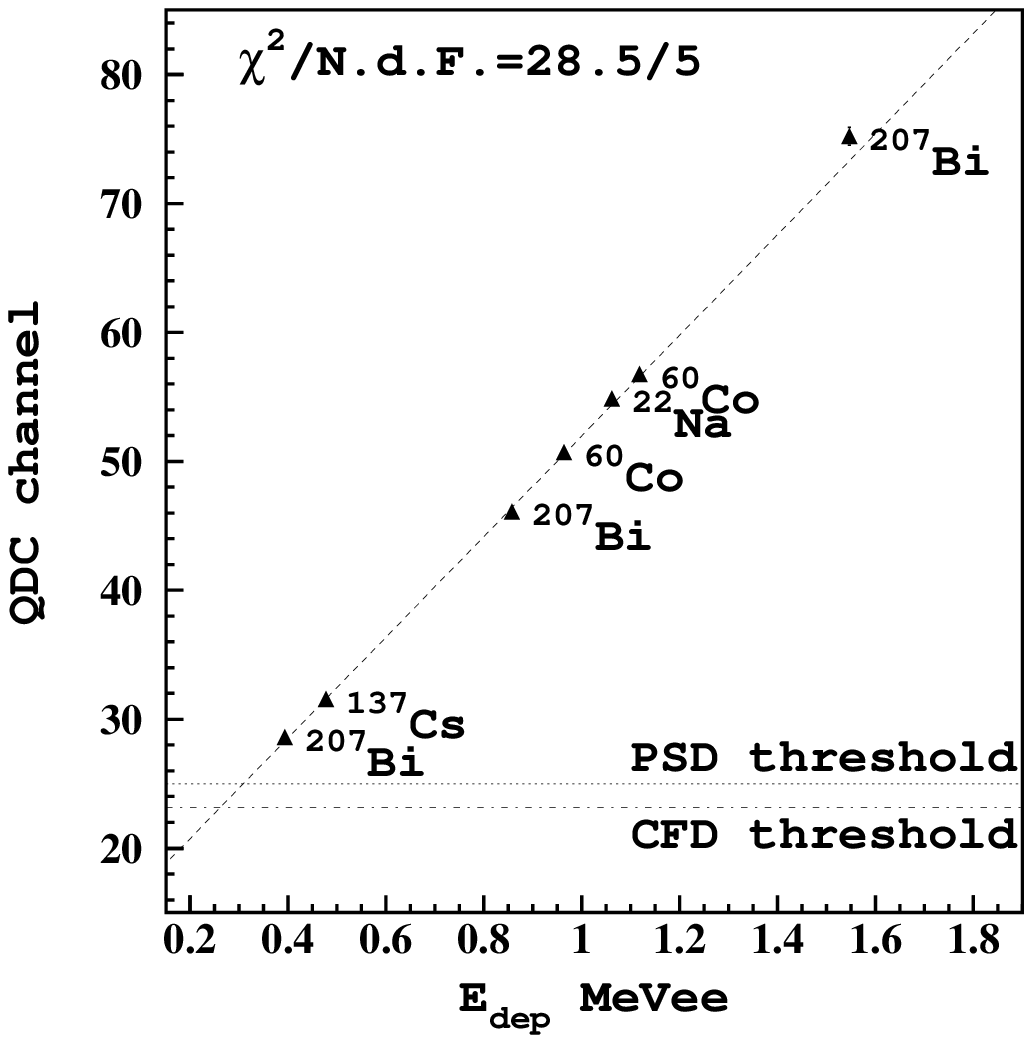}
\includegraphics[bb=2cm 6.5cm 20cm 24cm, scale=0.35]{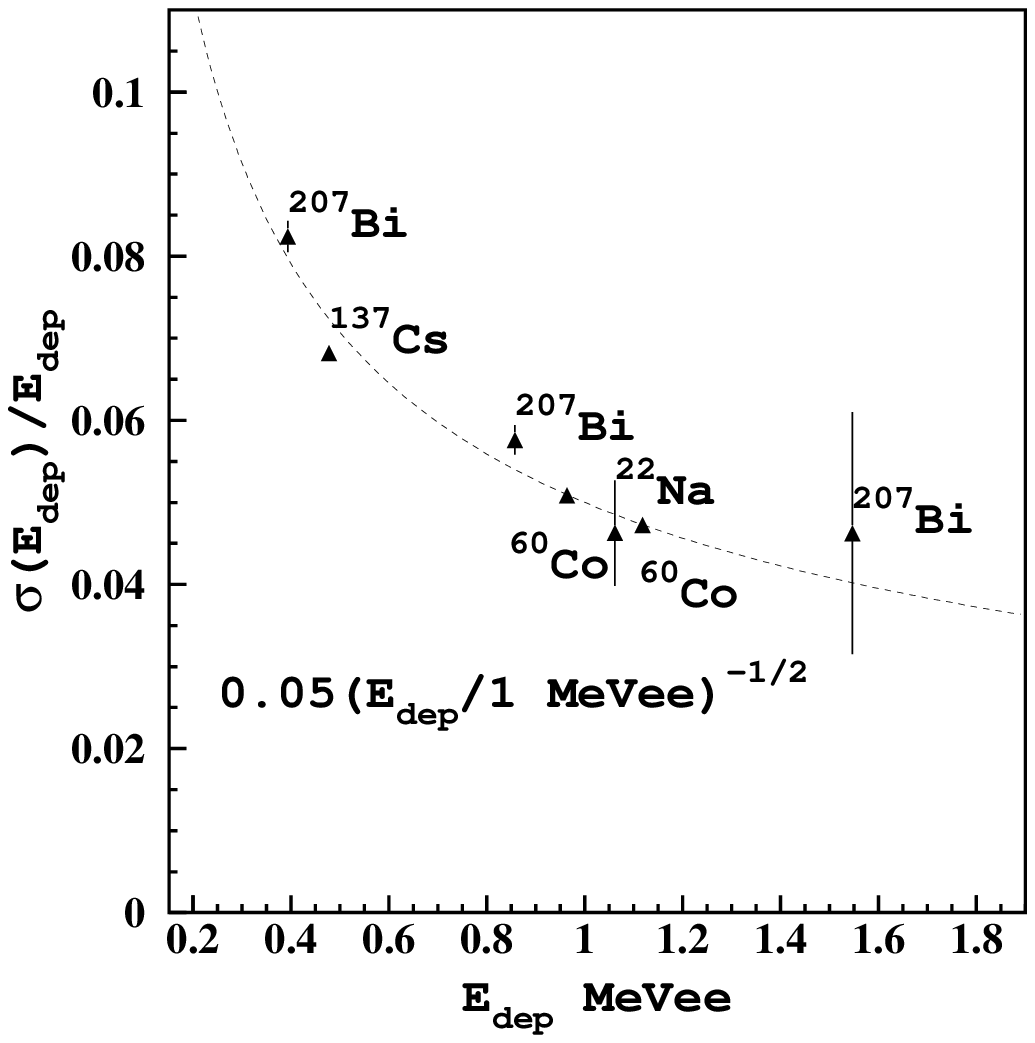}
\caption{\label{fig:nd1_qdc_calib}The same as in Fig.~\ref{fig:nd6_qdc_calib}, but for large
detectors.}
\end{center}
\end{figure}

Examples of obtained calibration fits for small and large detectors are shown in Figs.~\ref{fig:nd6_qdc_calib} and~\ref{fig:nd1_qdc_calib}. The employed set of calibration sources allowed to extend
the calibration data almost to the threshold.
The resolution of all detectors was found to be better than 5\% at 1 MeVee.

\subsection{Efficiency evaluation}\label{sec:effic}
The efficiencies of neutron detectors were obtained by means of Geant4.9.5-p01
Monte Carlo~\cite{geant4} simulations and verified in a dedicated measurement. 
The complete detector geometry was implemented into the simulation. Most of the materials were
obtained from the Geant4 NIST database. Electromagnetic physics list was
taken from G4EmStandardPhysics-option3 and G4EmExtraPhysics. Hadronic
and optical physics were also activated.
The cross sections used in Geant4 to describe the interaction of neutrons with
hydrogen and carbon were checked in the full neutron energy range of our data
and found to be in agreement with ENDF VII and Exfor databases
and with the data compilation from Ref.~\cite{DelGuerra}.
The scintillator light yield,
optical parameters of scintillator, reflector painting, glass and optical grease
were all configured accordingly. In particular, the reflector painting
was described as dielectric-dielectric optical surface using the unified
model and ground-front-painted finish with $\sigma_\alpha = 10$.

\begin{figure}[!ht]
\begin{center}
\includegraphics[bb=2cm 6.5cm 20cm 24cm, scale=0.35]{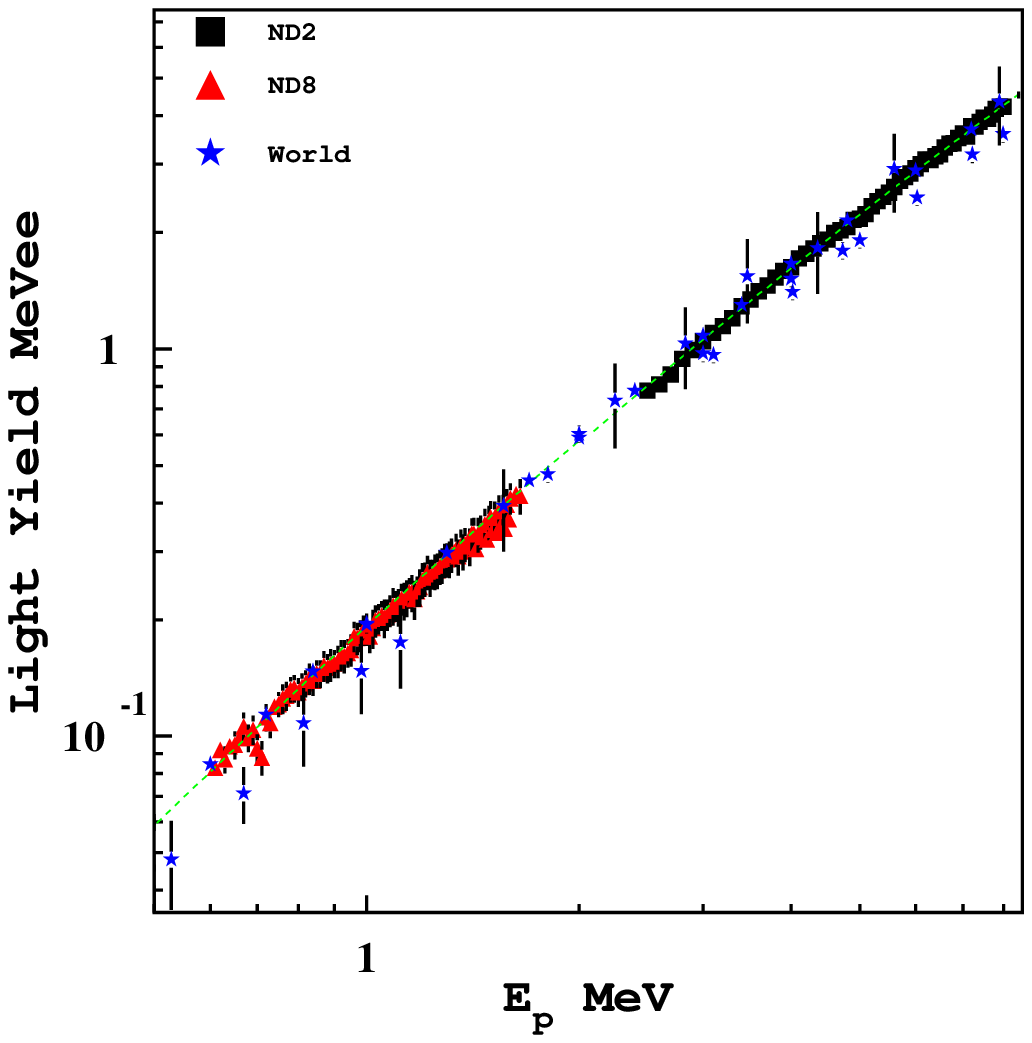}
\caption{\label{fig:fit_light_yield}(color online) Fit of world and this experiment data
on the light yield in the liquid scintillator.}
\end{center}
\end{figure}

A uniform neutron spectrum was generated from 0 to 62 MeV.
The detector efficiencies were obtained as the ratio of
detector hits depositing a visible energy above the experimental threshold
over the number of generated neutrons.
Since the evaluation relies on the visible energy, a careful check
of electromagnetic energy deposition and its saturation was performed.
The comparison of the standard Geant4 deposited energy in liquid scintillator
to SRIM2012 calculations~\cite{SRIM2012} showed significant disagreement for various particles.
We assumed that SRIM2012 calculations were more reliable and implemented
its energy loss tables for all relevant particles into Geant4.
The scintillation light quenching was activated with Birks parameter $k_B=0.007$ g/MeV/cm$^2$
obtained from a combined fit of all existing world data and the data from the present experiment
as shown in Fig.~\ref{fig:fit_light_yield}.
The introduction of a second parameter leads to a much better fit quality,
in particular in the low energy region.
However, a quadratic term like in Ref.~\cite{chou:quenching} introduces an unphysical ``unquenching''
behavior when extrapolated to high ionization density of $^{12}C$ ions.
A better description of the low-energy region was obtained with the following parametrization:
\begin{equation}
\Delta L=\frac{e^{-C (\frac{dE}{dx})^2}}{1+k_B\frac{dE}{dx}}\Delta E ~,
\end{equation}
\noindent with parameter $C=0.8\times10^{-6}$ (g/MeV/cm$^2$)$^2$.

The simulated efficiency below 10 MeV has been validated in a dedicated measurement, performed with the same experimental setup used for detector calibrations. The technique developed in Refs.~\cite{Lanzano,Colonna} was employed.
This consists of a coincidence measurement of the neutron and $\gamma$ from the same fission of a $^{252}$Cf source.
The same threshold values used in the experiment for large and small cells have been used.
In order to remove the background due to neutron rescattering from surrounding
materials the number of coincidences measured with the shadow bar,
consisting of a 50 cm long 4.6 cm diameter steel cylinder, installed in front
of the liquid scintillator detector was subtracted.

\begin{figure}[!ht]
\begin{center}
\includegraphics[bb=2cm 6.5cm 20cm 23cm, scale=0.35]{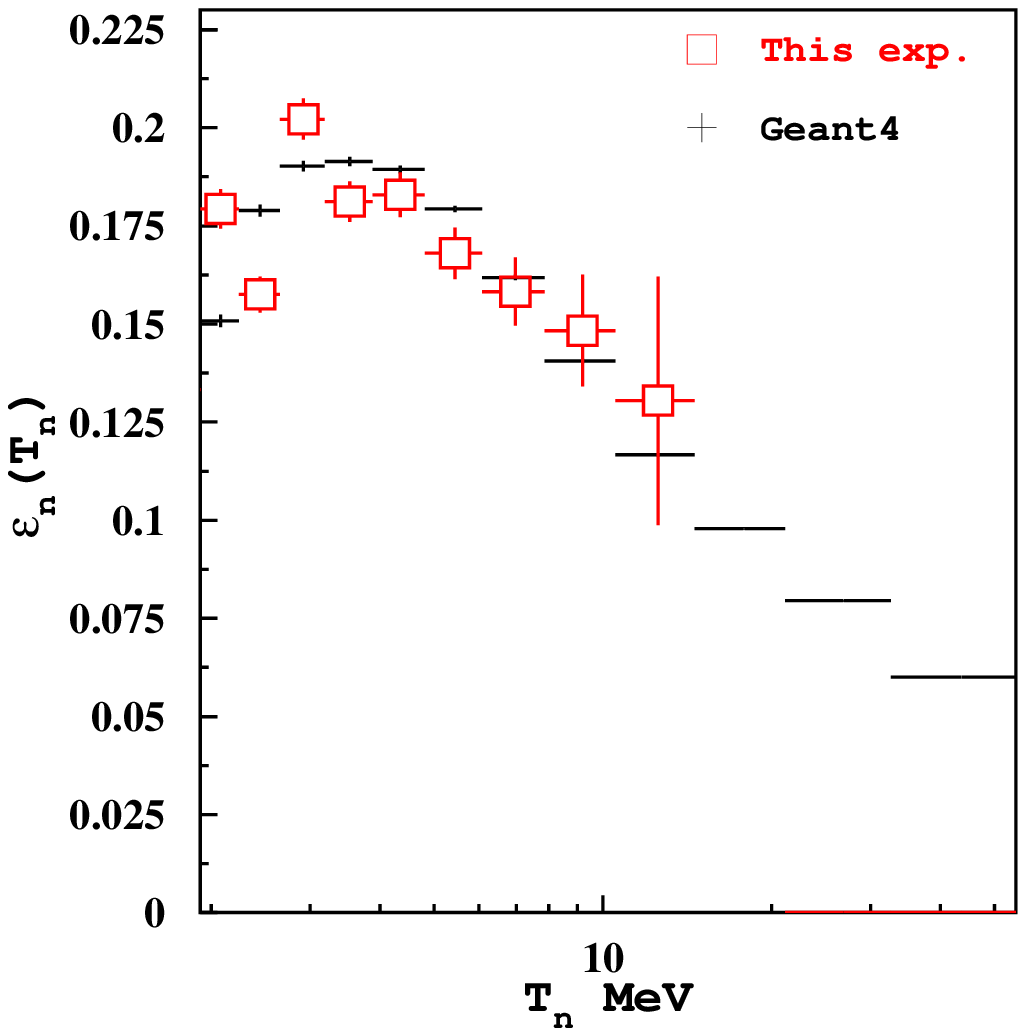}
\caption{\label{fig:g4_effs_large}(color online) Measured and simulated efficiency of large
liquid scintillator detector with the threshold of 263 keVee.}
\end{center}
\end{figure}

\begin{figure}[!ht]
\begin{center}
\includegraphics[bb=2cm 6.5cm 20cm 23cm, scale=0.35]{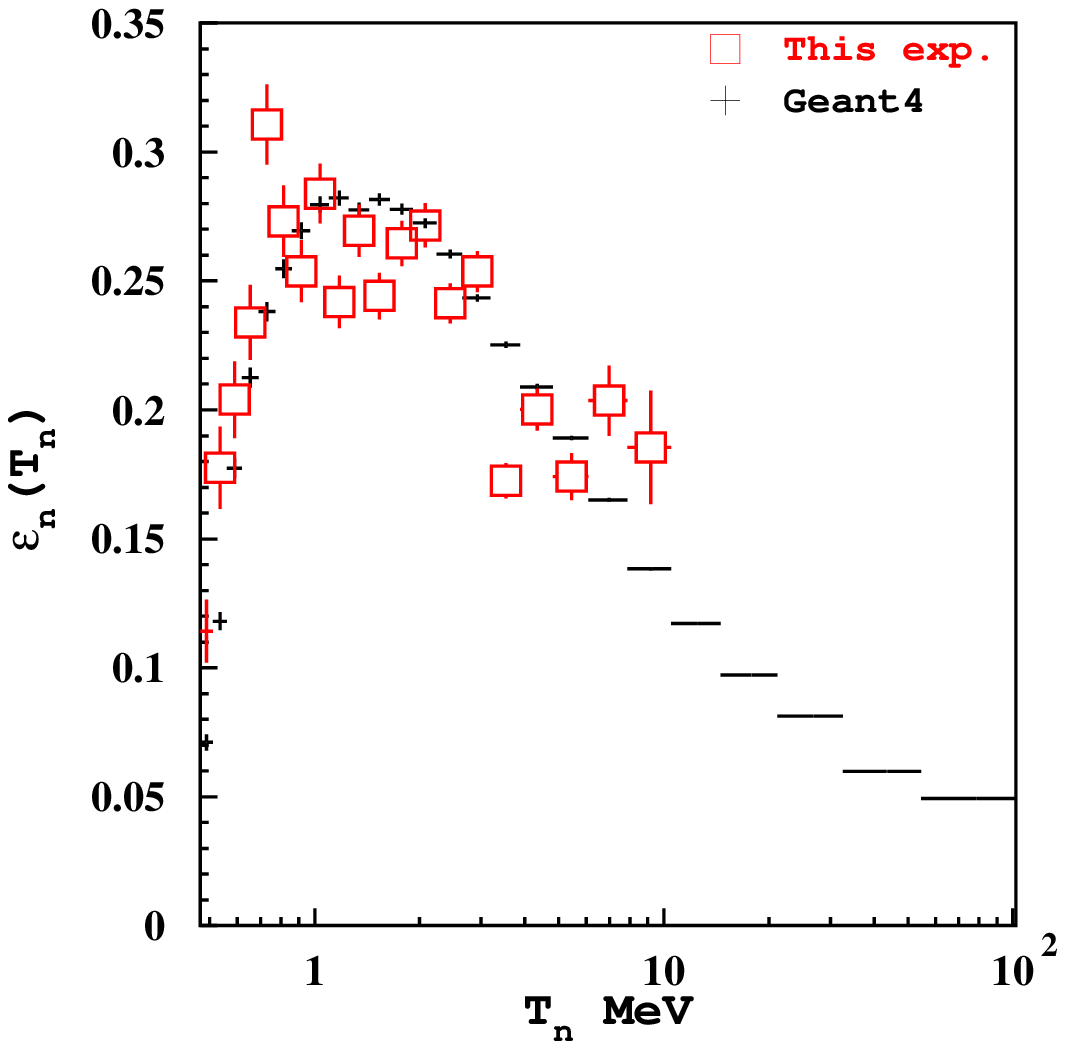}
\caption{\label{fig:g4_effs_small}(color online) The same as in Fig.~\ref{fig:g4_effs_large} but for small detector with the threshold of 54 keVee.}
\end{center}
\end{figure}

The obtained efficiencies for large and small detectors are shown in Fig.~\ref{fig:g4_effs_large}
and~\ref{fig:g4_effs_small}, respectively.
In the same figures, the simulated efficiencies in the full energy range useful for the data analysis are shown.
The measured efficiency was found to be in good agreement with Geant4 simulations over the entire energy range accessible with the $^{252}$Cf source.
However, an enhancement of the measured efficiencies with respect to the simulated ones was observed close to the detection threshold. We supposed that it was due to some residual background, visible here because of the very low number of measured events. This hypothesis has been verified with a dedicated simulation in which neutrons emitted by the source were allowed to be scattered by the BaF$_2$ detector and by its support. The absolute number and the energy distribution of the scattered neutrons entering the liquid scintillator along the path covered by the shadow bar during the background measurement, and therefore not accounted for in the background subtraction, were found to be compatible with the observed discrepancies.

These results together with the check performed on the neutron interaction cross sections used in Geant4
make us confident about using the Geant4 simulated efficiency.


In order to avoid large systematic uncertainties only the energy region
where the efficiency exceeded 3\% was selected for the data analysis.

\section{Data analysis}
\subsection{Event selection}
At each angular setting the data were taken in few runs. In order to check the consistency
of all the data and exclude runs with significant variations in detector conditions,
a set of quality checks were performed.
These checks include:
\begin{enumerate}
\item TDC and QDC pedestals,
\item total number of triggers normalized to the collected beam charge,
\item prompt photon peak position.
\end{enumerate}
Pedestals of TDC and QDC were stable except for some runs of large detectors
installed at 0 and 5 degrees.

The total number of triggers normalized to the collected beam charge
varied from 0.1\% to 13\% with average of 2.5\%.
We selected only those runs where the variation was below 5\%.

The stability of the prompt photon peak position was verified over
each set of runs. The runs where the prompt photon peak position resulted
to be different from the average were discarded.

Two independent DAQ systems were used for small and large detectors.
Each DAQ stored the data from four detectors, installed at different angles.
Identification of trigger channel was performed during the off-line analysis.
The first detector hit, corresponding to largest TDC value above pedestal,
identified the trigger channel.
The probability of simultaneous hits in more
than one detector working at a few kHz overall rate within the 250 ns time interval covered by acquisition Gate was negligible.

\subsection{Particle identification}
\begin{figure}[!ht]
\begin{center}
\includegraphics[bb=2cm 6.5cm 20cm 24cm, scale=0.35]{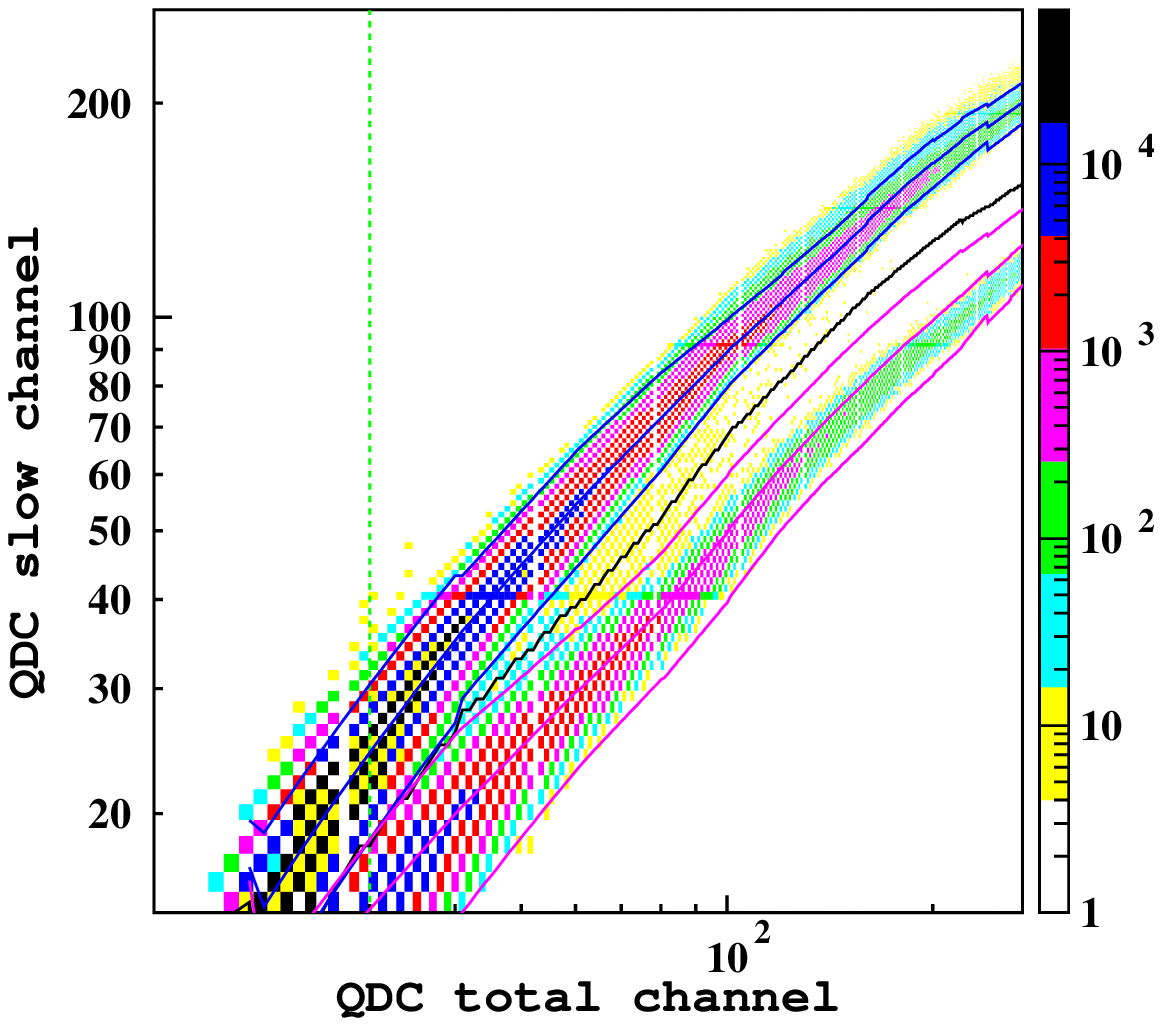}
\caption{\label{fig:psd_cuts_nd4}(color online) PSD distribution in large detector.
The green dashed line shows the minimal PSD threshold;
the magenta curves are fitted $\gamma$ peak position and width ($\pm3\sigma$);
the blue curves are neutron peak position and width;
the black curve represents the cut between $\gamma$s and neutron minimizing the sum
of inefficiency and relative contamination.}
\end{center}
\end{figure}

\begin{figure}[!ht]
\begin{center}
\includegraphics[bb=2cm 6.5cm 20cm 24cm, scale=0.35]{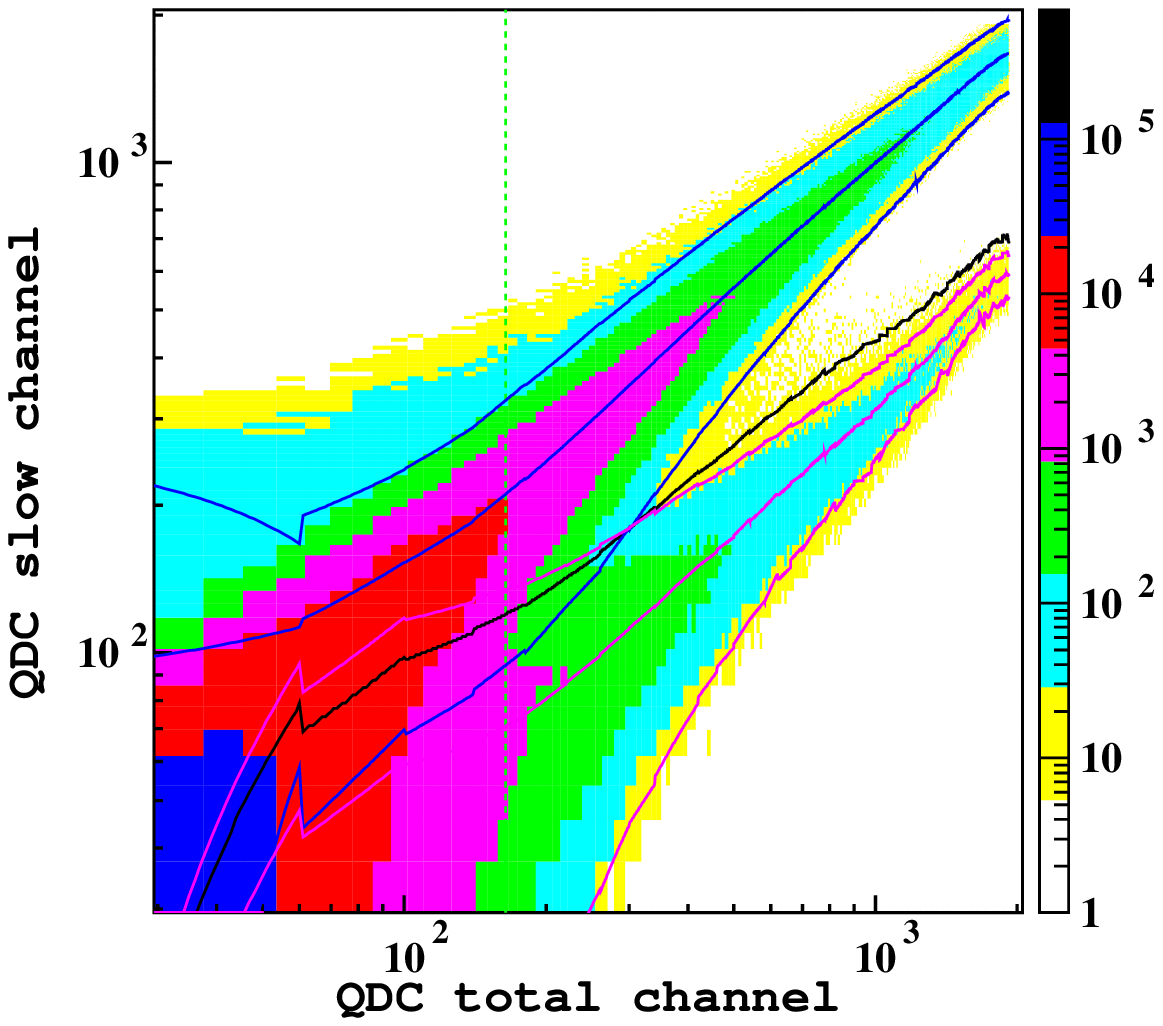}
\caption{\label{fig:psd_cuts_nd6}(color online) The same as in Fig.~\ref{fig:psd_cuts_nd4},
but for small detectors.}
\end{center}
\end{figure}

Neutron hits in the detectors were identified within the data
by a series of conditions. These included the following requirements:
\begin{enumerate}
\item the hit time with respect to the RF signal of the beam had to be smaller than the RF period of 125 ns,
moreover also the amplitude of TFC signal was tuned to cover only the measurable range;
\item the hit time had to be larger than the minimum time necessary to a neutron of 62 MeV
to travel the distance from the target to detector minus 3$\sigma$ of timing resolution,
obtained by fitting the prompt photon peak;
\item the electron equivalent energy deposited in the liquid scintillator had to be greater than zero and smaller than the maximum electron equivalent energy produced by interaction of a neutron with measured Time-of-Flight (kinetic energy) $E_{dep}^{max}(T_n)$.
\item deposited energy had to be greater than PSD applicability threshold.
The thresholds were determined by the condition that both neutron loss and $\gamma$ contamination
were below 1\%.
The PSD thresholds were significantly higher than the hardware (discriminator) thresholds
and varied from 44 to 54 keVee for small detectors and from 175 to 308 keVee for large detectors.
\item the integrated slow scintillation component has to be higher
than the fitted PSD separation curve.
The determination of the optimal cut on the contribution of the signal tail
in the total was obtained by fitting the PSD distributions for $\gamma$s and neutrons
shown in Figs.~\ref{fig:psd_cuts_nd4} and~\ref{fig:psd_cuts_nd6}. To this end the data taken without beam
were analyzed in order to obtain $\gamma$ PSD distribution. This procedure was repeated using
the data taken with beam on, by fitting both $\gamma$ and neutron distributions simultaneously.
\end{enumerate}

\begin{figure}[!ht]
\begin{center}
\includegraphics[bb=2cm 6.5cm 20cm 23cm, scale=0.35]{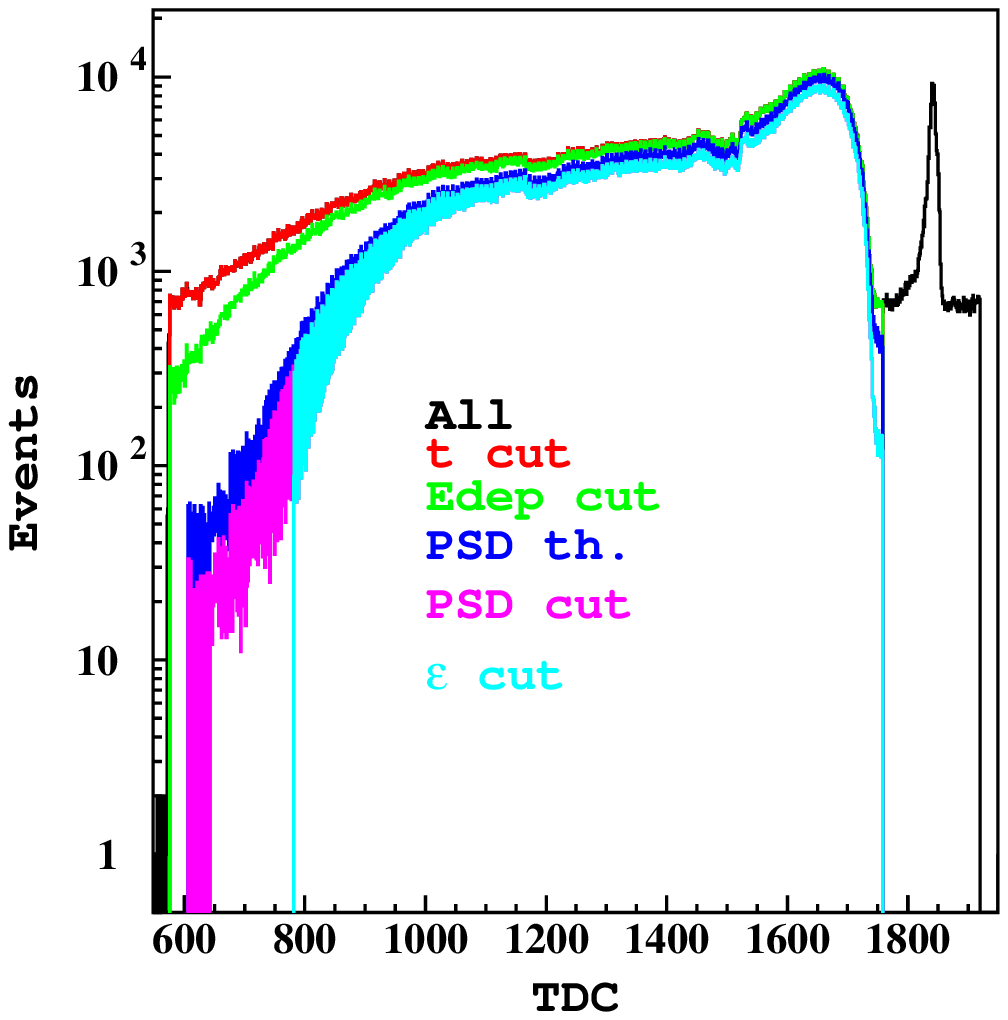}
\includegraphics[bb=2cm 6.5cm 20cm 24cm, scale=0.35]{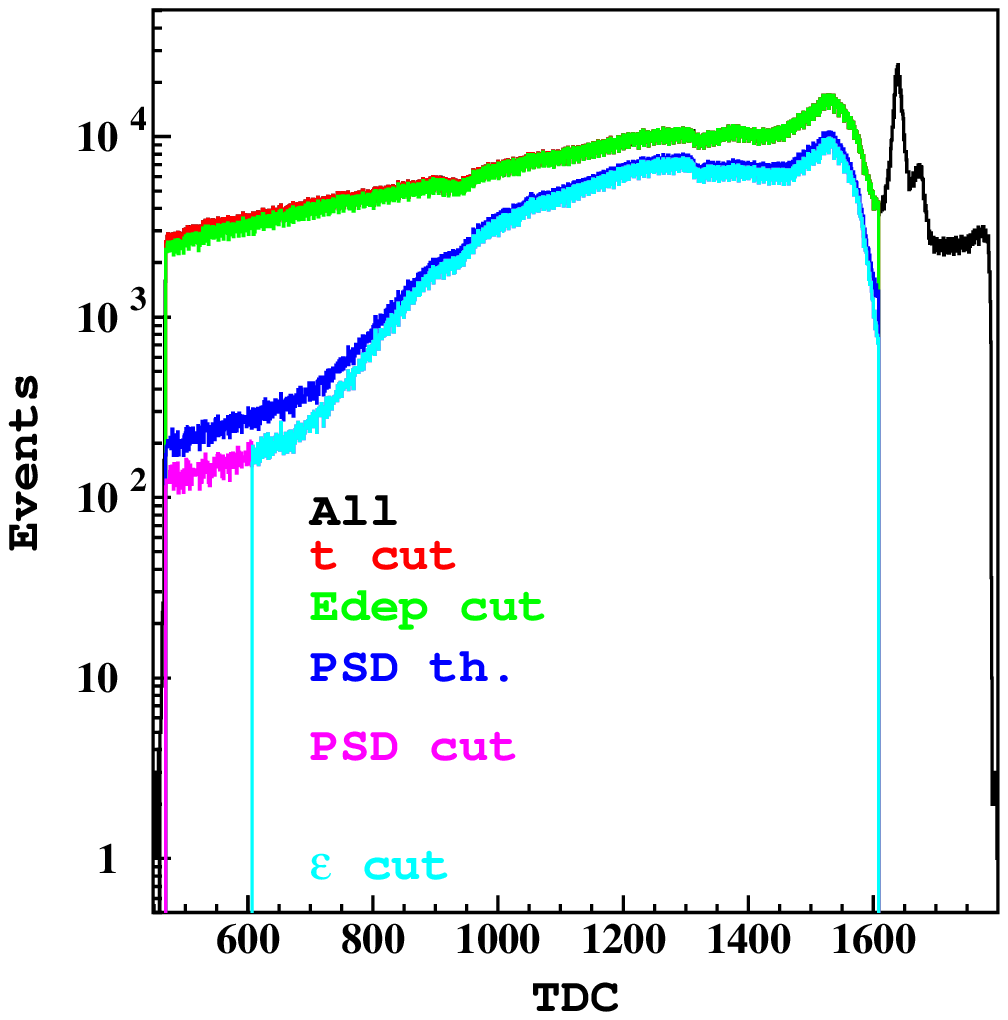}
\caption{\label{fig:cuts_pid}(color online) Effect of different PID cuts on the TDC
distributions measured with large detector at 30 degrees (upper) and small detector at 60 degrees (lower):
black - all the data;
red - after timing cuts;
green - after timing and deposited energy cuts;
blue - after timing, deposited energy and PSD threshold cuts;
magenta - after timing, deposited energy, PSD threshold and PSD cuts;
cyan - after timing, deposited energy, PSD threshold, PSD and minimal efficiency cuts.}
\end{center}
\end{figure}

Examples of contribution of each of these conditions are shown in Fig.~\ref{fig:cuts_pid}.
The timing cuts mainly remove the prompt photon peak. The deposited energy cut
reduces the event yield at small TDC values corresponding to low energy neutrons.
This is because background events usually have deposited energy larger than
allowed for a neutron of such small kinetic energy.
PSD threshold cut leads to the most severe reduction of events and to the corresponding
loss of efficiency, in particular in the low neutron energy range.
PSD cut removes the remaining $\gamma$ background, as it can be seen from the strong reduction
of events just before the prompt photon peak.
The minimal efficiency cut simply limits the measurable TDC range.

\subsection{Collected Beam Charge Normalization}\label{sec:fc_charge}
\begin{figure}[!ht]
\begin{center}
\includegraphics[bb=2cm 6cm 20cm 23cm, scale=0.35]{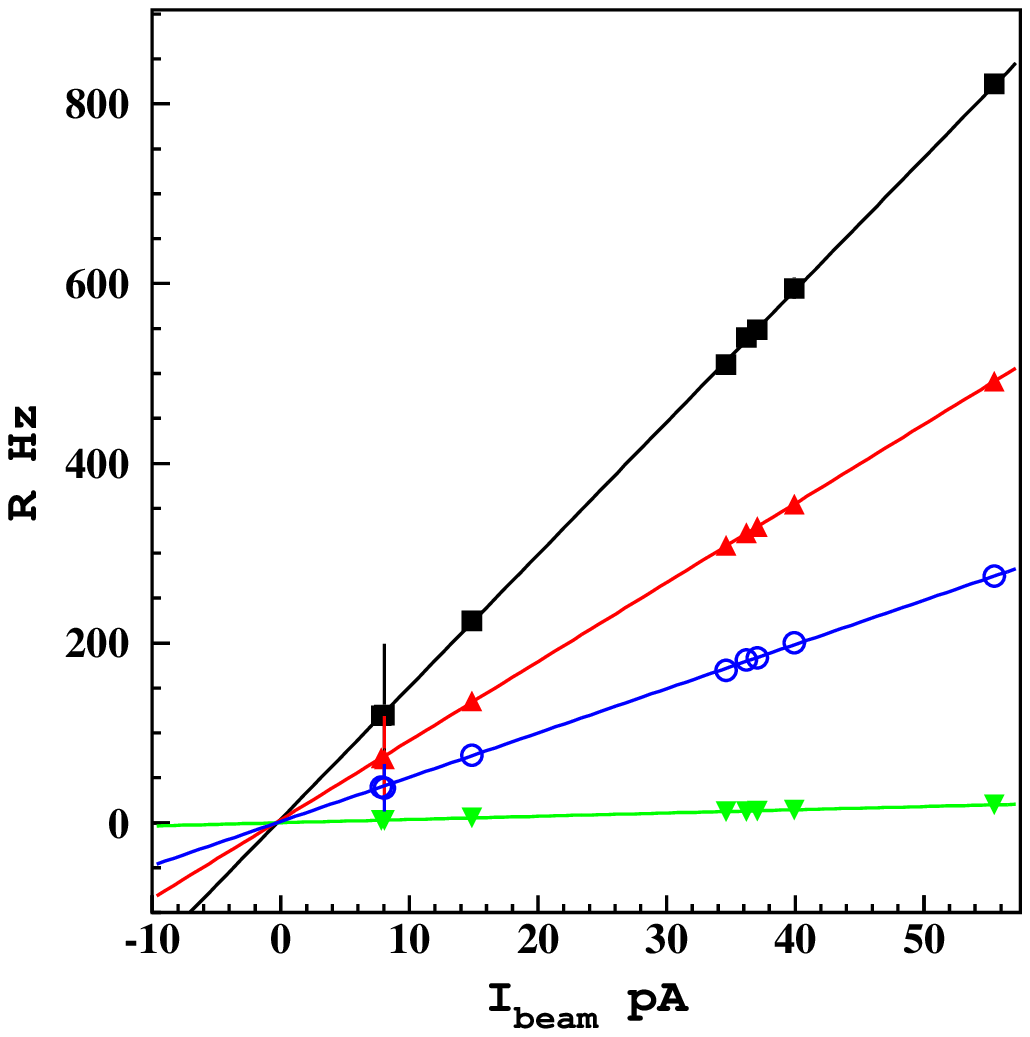}
\caption{\label{fig:ramp_fcq}(color online) Total event rate dependence on the measured beam current.
Different colors indicate four independent detectors installed at different
angles. Straight lines show linear fits to the data. The extrapolated beam current
at the intercept with $R=0$ axis (offset current) is compatible with zero.}
\end{center}
\end{figure}

The charge deposited by the proton beam on the Beryllium target was read out
via direct electrical contact using a high precision current digitizer.
During the experiment two different current digitizers were employed:
model 1000C of Brookhaven Instruments Corporation with Low Current Modification
and Ortec model 439.
Test bench measurements showed that the precision of these two modules
is better than 2 pA. We consider this value as a constant contribution to
the normalization systematic uncertainty.
Cross checks of measured charges were performed exploiting
the linear dependence of the total event rate on the beam current by extrapolating the
the beam current value to zero event rate to obtain an estimate of the current offset.
An example of such cross check measurement is shown in Fig.~\ref{fig:ramp_fcq}.
The offset currents obtained in this way, ranging from -3 to 0 pA (Ortec 439) and to 6.4 pA (BIC 1000C),
were used to correct the final data.
The values of this correction were taken as the normalization systematic uncertainty.

From the measured total beam charge deposited on the target $Q_{beam}=\int_{t_{run}} I_{beam} (t) dt$
the total number of protons on target was obtained. This was performed correcting the collected charge
for the DAQ live-time fraction $L_{DAQ}$:
\begin{equation}
N_p=\frac{Q_{beam} L_{DAQ}}{e}~,
\end{equation}
\noindent where $e$ is the elementary charge and $L_{DAQ}$ is the live-time correction. The latter was obtained in each run as the ratio between the number of recorded events and the number of trigger events measured
by the scaler, corrected for multiple triggers.
The measured DAQ dead time of 13 $\mu$s resulted to be in agreement with expectations.

\subsection{Background subtraction}
\begin{figure}[!ht]
\begin{center}
\includegraphics[bb=2cm 6.5cm 20cm 23cm, scale=0.35]{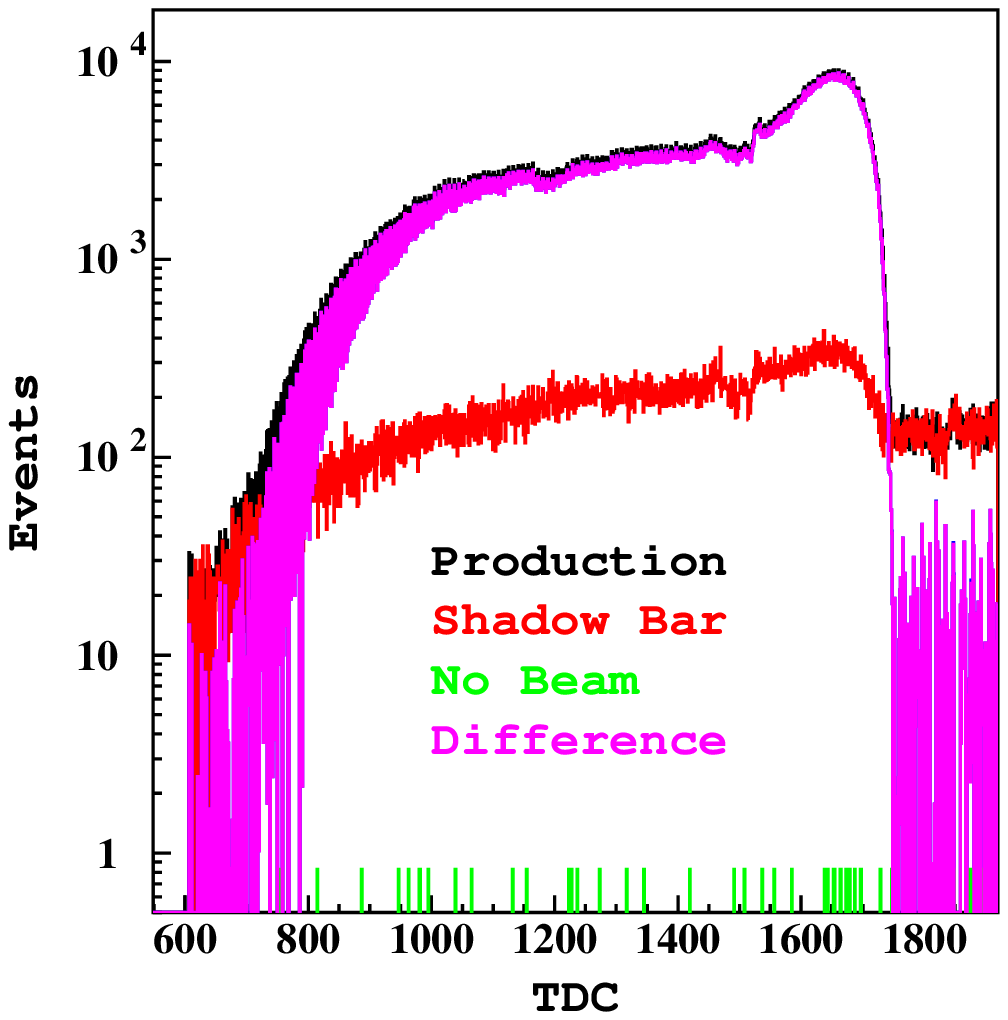}
\includegraphics[bb=2cm 6.5cm 20cm 24cm, scale=0.35]{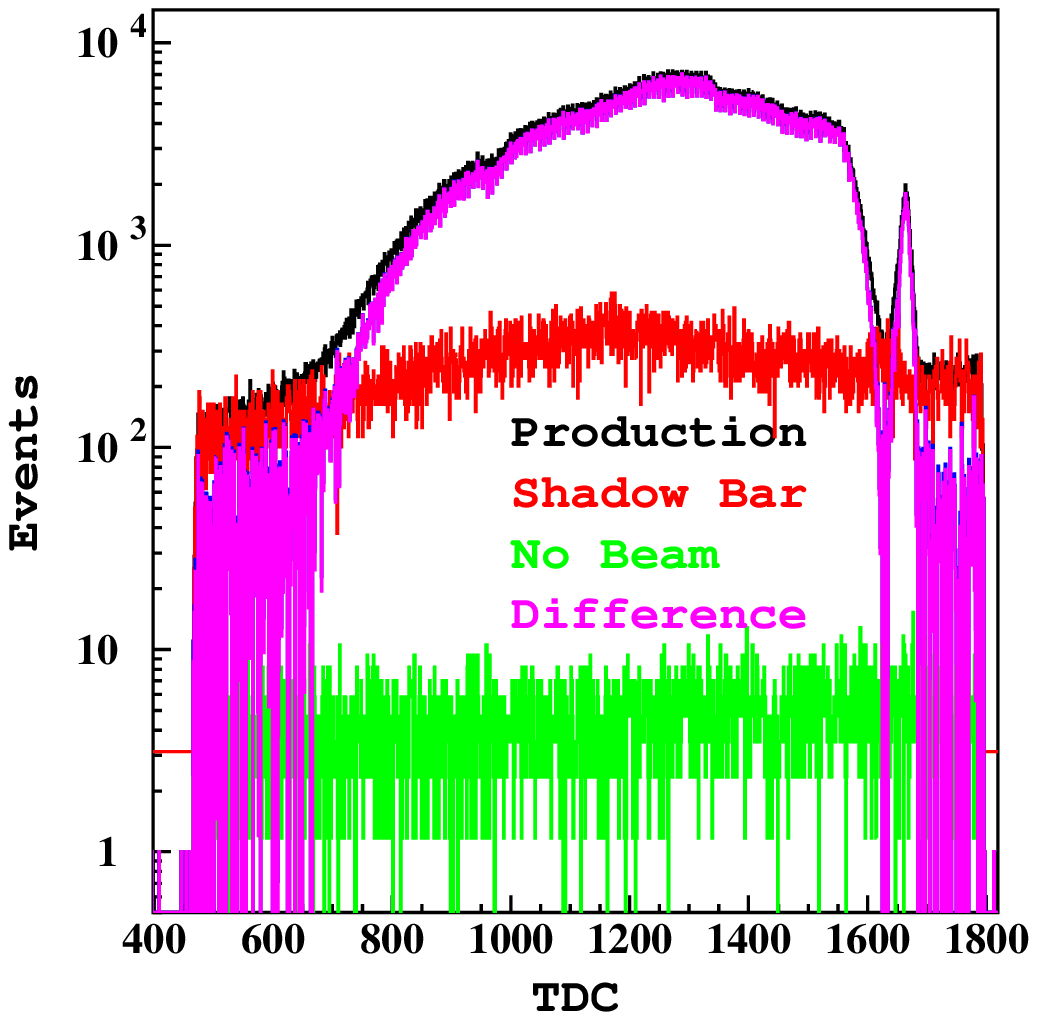}
\caption{\label{fig:bkg_subtract}(color online) Shadow bar background and noise subtraction on the TDC
distributions measured with a large detector at 30 degrees (upper) and a small detector at 90 degrees (lower):
All cuts were applied to the data except for the timing conditions.
The histograms show the following normalized spectra:
black - production data;
red - data taken with shadow bar obscuring the target view;
green - noise measured with beam off;
magenta - after background subtraction.}
\end{center}
\end{figure}

Neutrons scattered from surrounding materials, like beamline or detector supports,
experimental hall floor or walls and other installations nearby, may eventually get into detectors.
These hits will be registered as good neutron signals if the deposited energy
will remain below $E_{dep}^{max}$. As mentioned before, in this experiment particular attention
was paid to reduce the material around the detectors.
Nevertheless, some fraction of events comes from environment-scattered neutrons.
In order to measure these environmental neutrons, special runs were taken
by screening the target direct view from the detector with a shadow bar.
The shadow bar consisted of a 50 cm long 4.6 cm diameter steel cylinder.
MCNP/Geant4 simulations indicated that such a shadow bar allows
to reduce the initial neutron flux in the energy range
from 0.5 to 62 MeV by a factor $<10^{-3}$.
Therefore, the data taken with shadow bar measured the contribution
of environmental neutrons.
Indeed, after subtraction of the shadow bar yields, for example shown in Fig.~\ref{fig:bkg_subtract},
the neutron yield outside the physical range (above prompt $\gamma$ peak) goes to zero.
The contribution of this background varies from a few percent in average to
$>50$ \% at the endpoints of measurable energy interval.

Another background was due to PMTs dark current noise. This was measured
during the runs without beam. For the large detectors this contribution
was negligible. However, for small detectors whose discriminator threshold
was set to fairly low value this background could reach a few percent.
The contribution of this background is shown in Fig.~\ref{fig:bkg_subtract}
in comparison with environmental background.

\subsection{Differential neutron yield}
The measured neutron spectra were divided in 30 variable energy bins
whose size was calculated according to the detector neutron energy resolution. The detector resolution
was estimated as a combination in quadrature of ToF indeterminations due to target thickness,
detector thickness, beam bunch width and PMT risetime. For each neutron energy the
bin size was taken to be equal to four times RMS value of the obtained resolution.

The differential neutron yields were extracted from the number of measured neutron hits
in the detector $N_n^{det.}(T_n,\theta_n)$ according to the following equation:
\begin{equation}
\frac{1}{N_p}\frac{d^2 N_n}{dT_n d\Omega_n}(T_n,\theta_n) =
\frac{1}{\epsilon_n(T_n) N_p}\frac{N_n^{det.}(T_n,\theta_n)}{\Delta T_n \Delta \Omega_n}~,
\end{equation}
\noindent where $N_p$ is the total number of incident protons on target,
$\epsilon_n(T_n)$ is the neutron detection efficiency obtained in Geant4 simulations,
$T_n$ and $\theta_n$ are neutron kinetic energy and production angle with respect to the beam axis, respectively. $\Delta T_n$ is the neutron energy bin size and $\Delta \Omega_n$ is the detector solid angle.

In order to cover the entire measured energy range we combined the data from
the small and large detectors. The maximum deposited energy measurable with the small
detectors was 0.5 MeVee, corresponding to maximum neutron energy of about 2 MeV.
Neutrons with energies above 2 MeV were also measured in small detectors,
but it was impossible to apply PSD and maximum deposited energy cuts for these events.
On the other hand, the efficiency of the large detectors dropped rapidly below 2 MeV,
leading to a large systematic uncertainty on the observed yield.
Therefore, we selected data from the small detector at $T_n <2$ MeV and large
detector data at $T_n >2$ MeV.
In the overlap region the two detector types show a fairly good agreement
with average deviation of the order of 10\%.

\begin{figure}[!ht]
\begin{center}
\includegraphics[bb=2cm 5.8cm 20cm 23cm, scale=0.35]{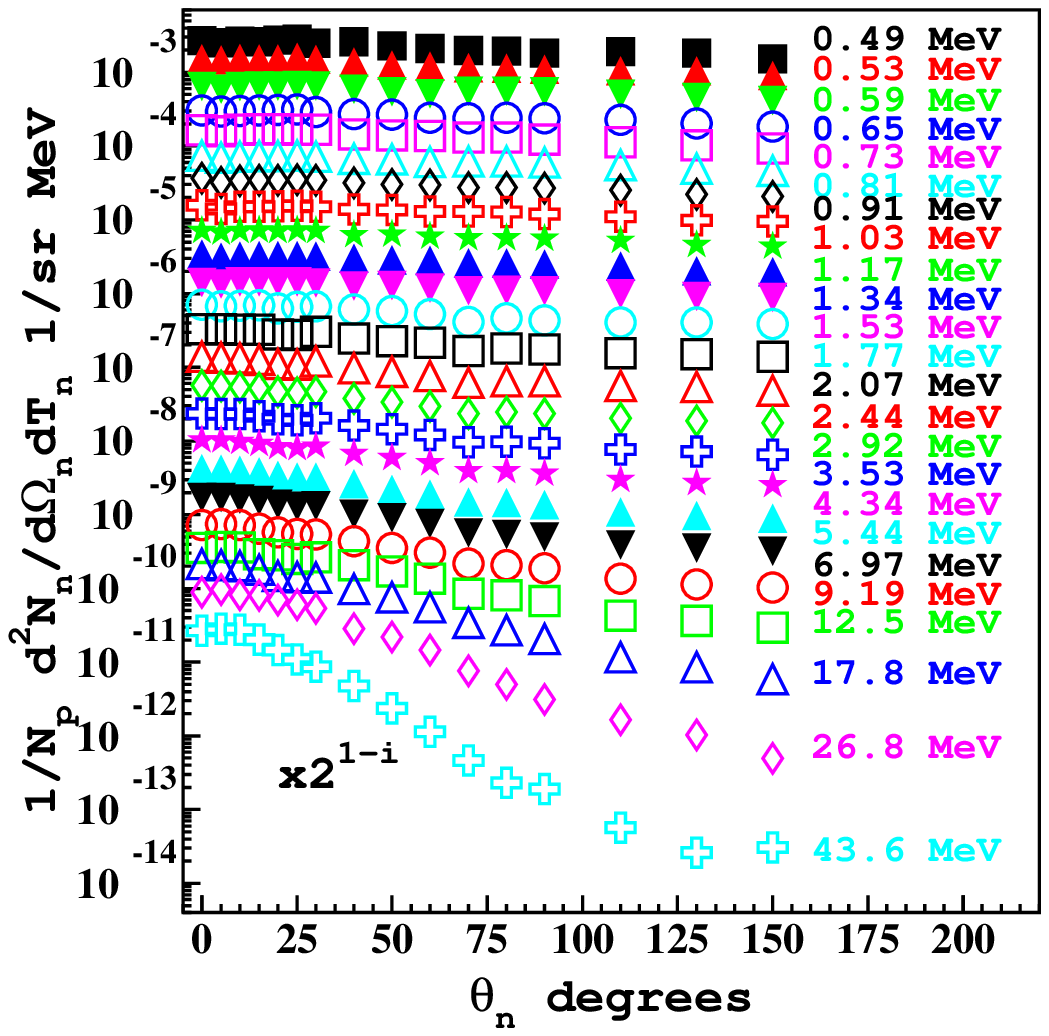}
\caption{\label{fig:ang_yield}(color online) Angular dependence of the neutron yield
for different neutron energies.}
\end{center}
\end{figure}

\begin{figure}[!ht]
\begin{center}
\includegraphics[bb=2cm 5.8cm 20cm 23cm, scale=0.35]{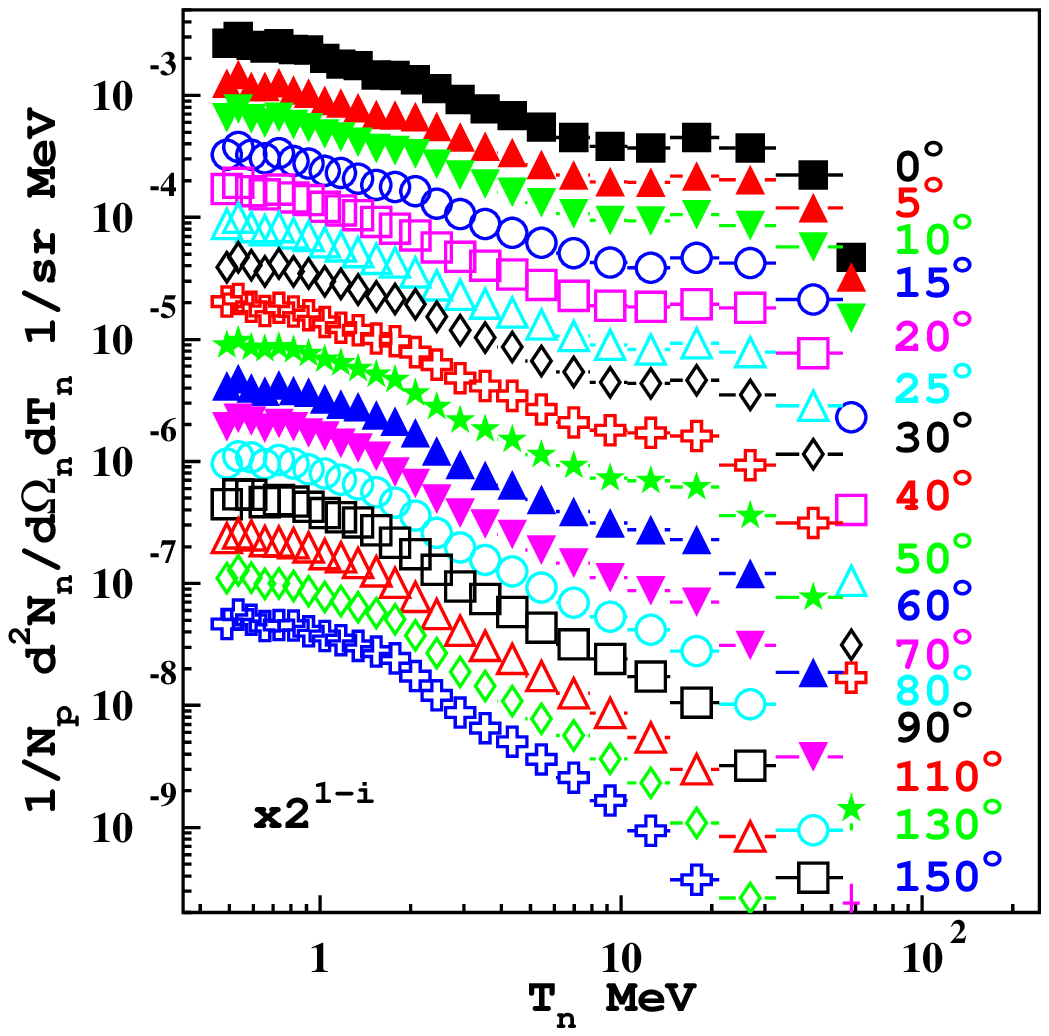}
\caption{\label{fig:eng_yield}(color online) Energy dependence of the neutron yield
for different production angles.}
\end{center}
\end{figure}

The obtained yields are shown in Figs.~\ref{fig:ang_yield} and~\ref{fig:eng_yield}
as a function of neutron angle and energy, respectively.
At low neutron energy the yield shows an isotropic behavior, while becoming forward-peaked at
for higher energy. The energy dependence is well described by exponential fall-off with a
high energy knee observed at $\theta_n<60^\circ$.

Statistical uncertainties on $N_n^{det.}$ and $\epsilon_n$ were combined in quadrature.
The two systematic uncertainties described in section~\ref{sec:sys_err} instead were kept
separate except for the fully integrated yield for which they were combined in quadrature.

\begin{figure*}[hpt]
\begin{center}
\includegraphics[bb=2cm 5.5cm 20cm 23cm, scale=0.32]{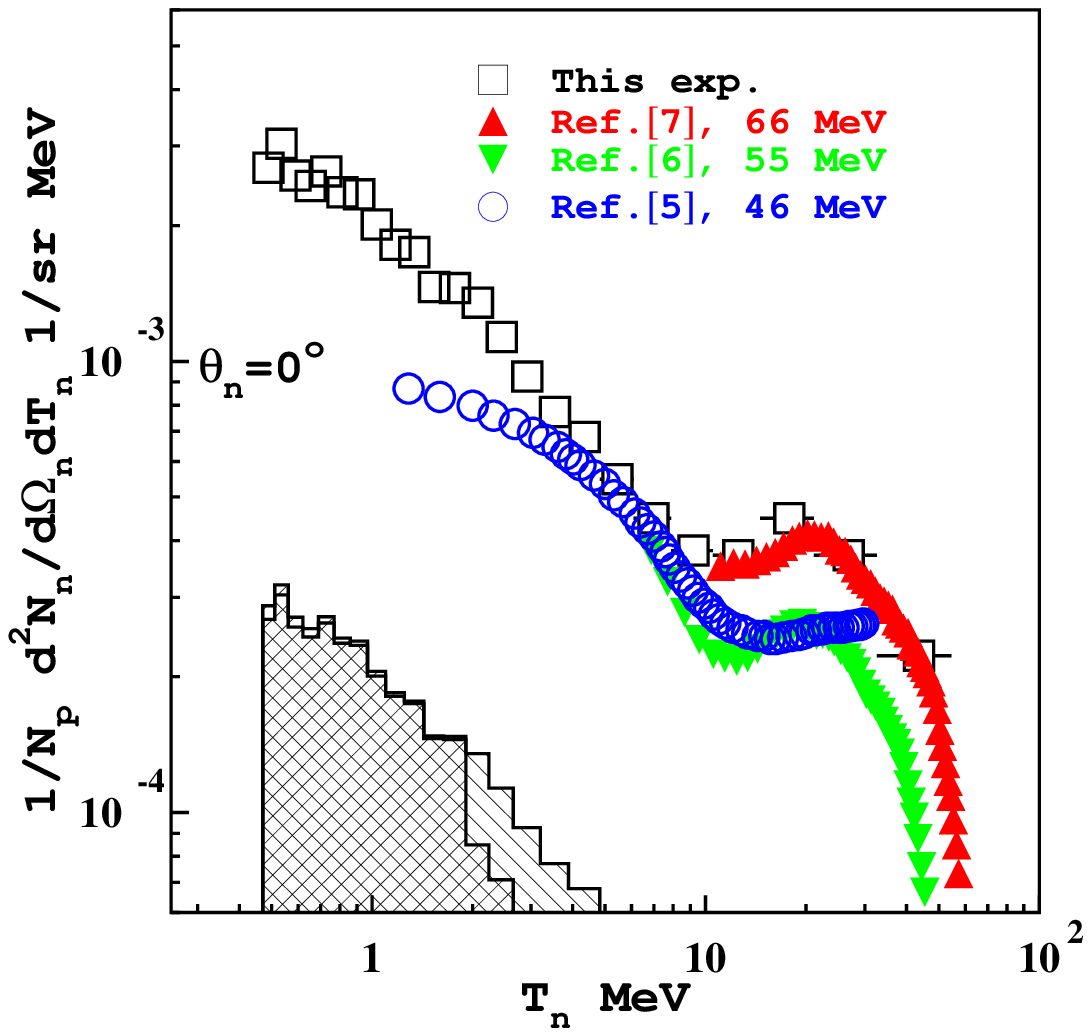}~%
\includegraphics[bb=2cm 5.5cm 20cm 23cm, scale=0.32]{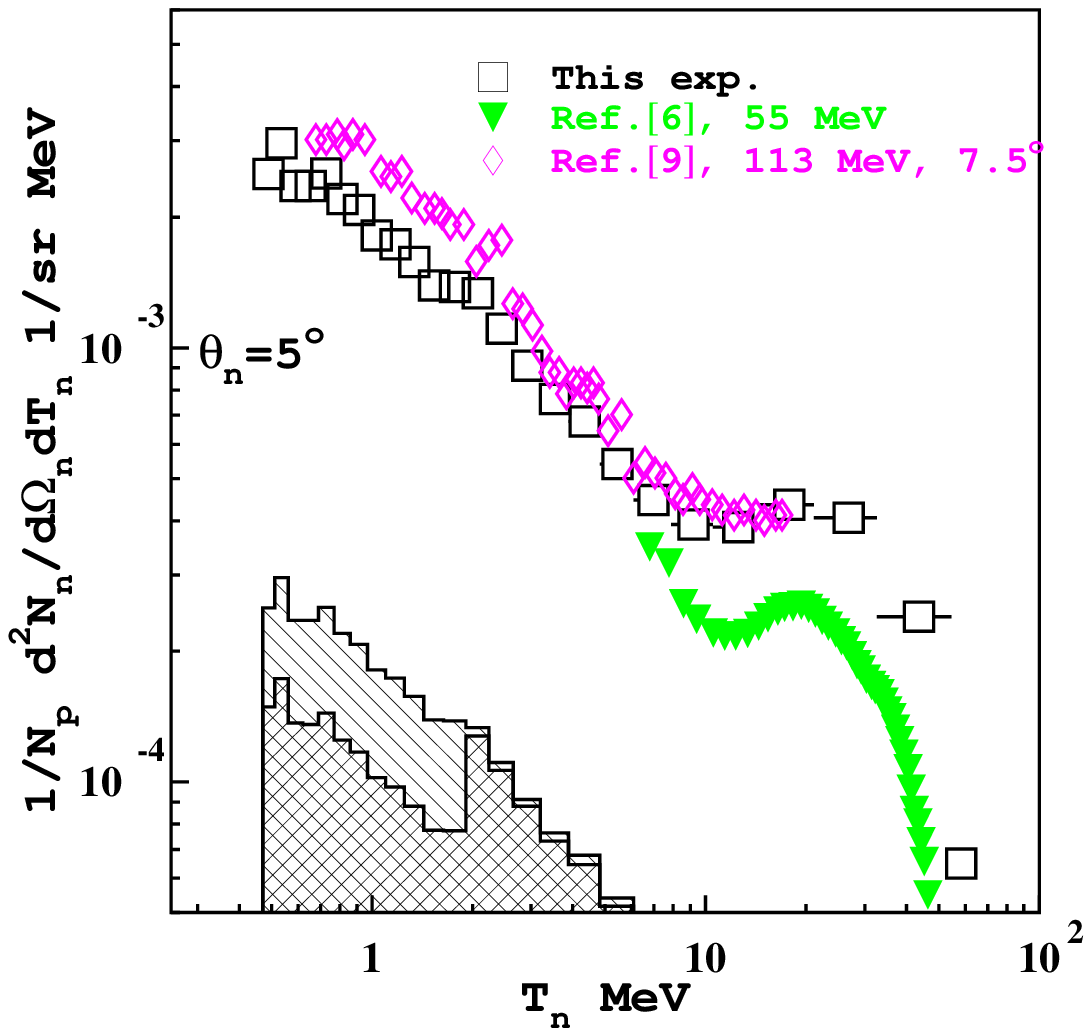}
\includegraphics[bb=2cm 5.5cm 20cm 23cm, scale=0.32]{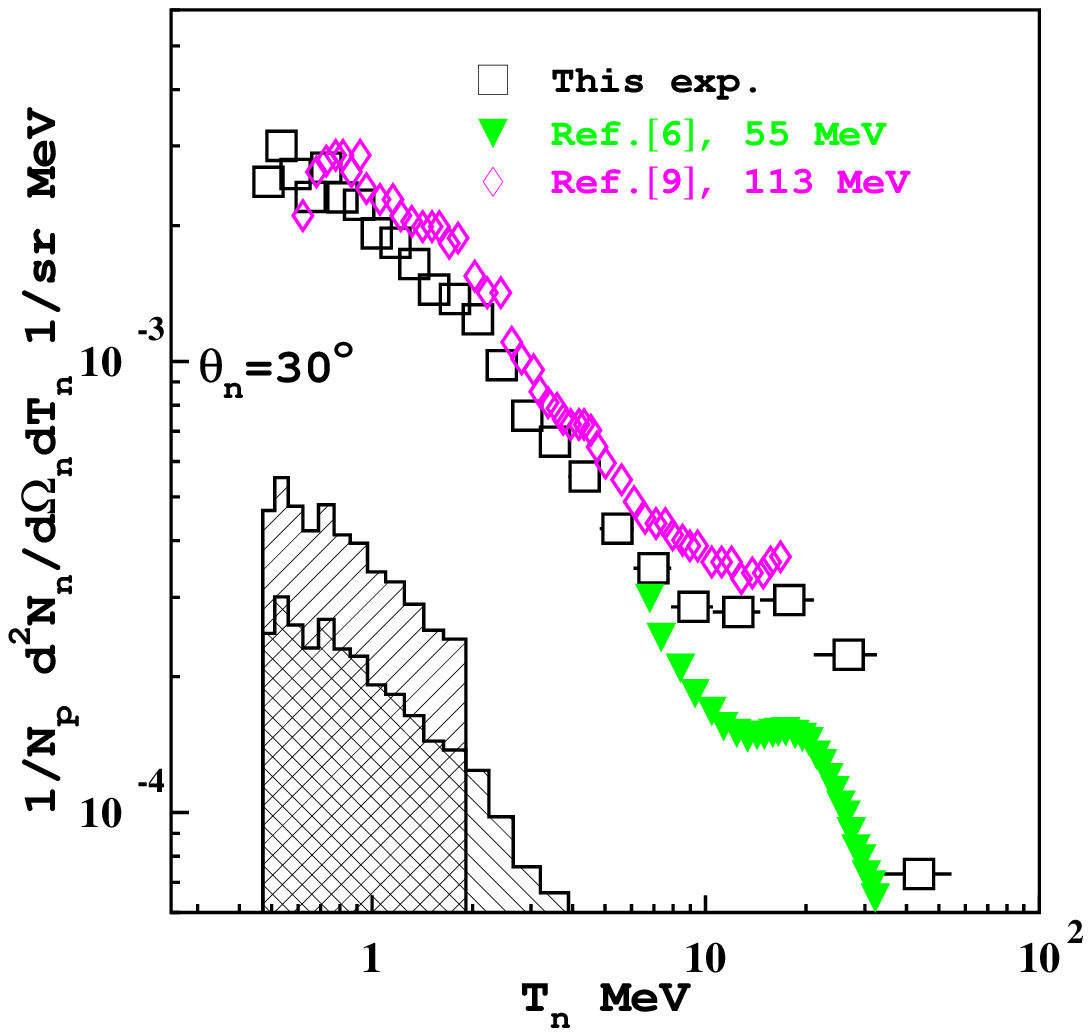}~%
\includegraphics[bb=2cm 5.5cm 20cm 23cm, scale=0.32]{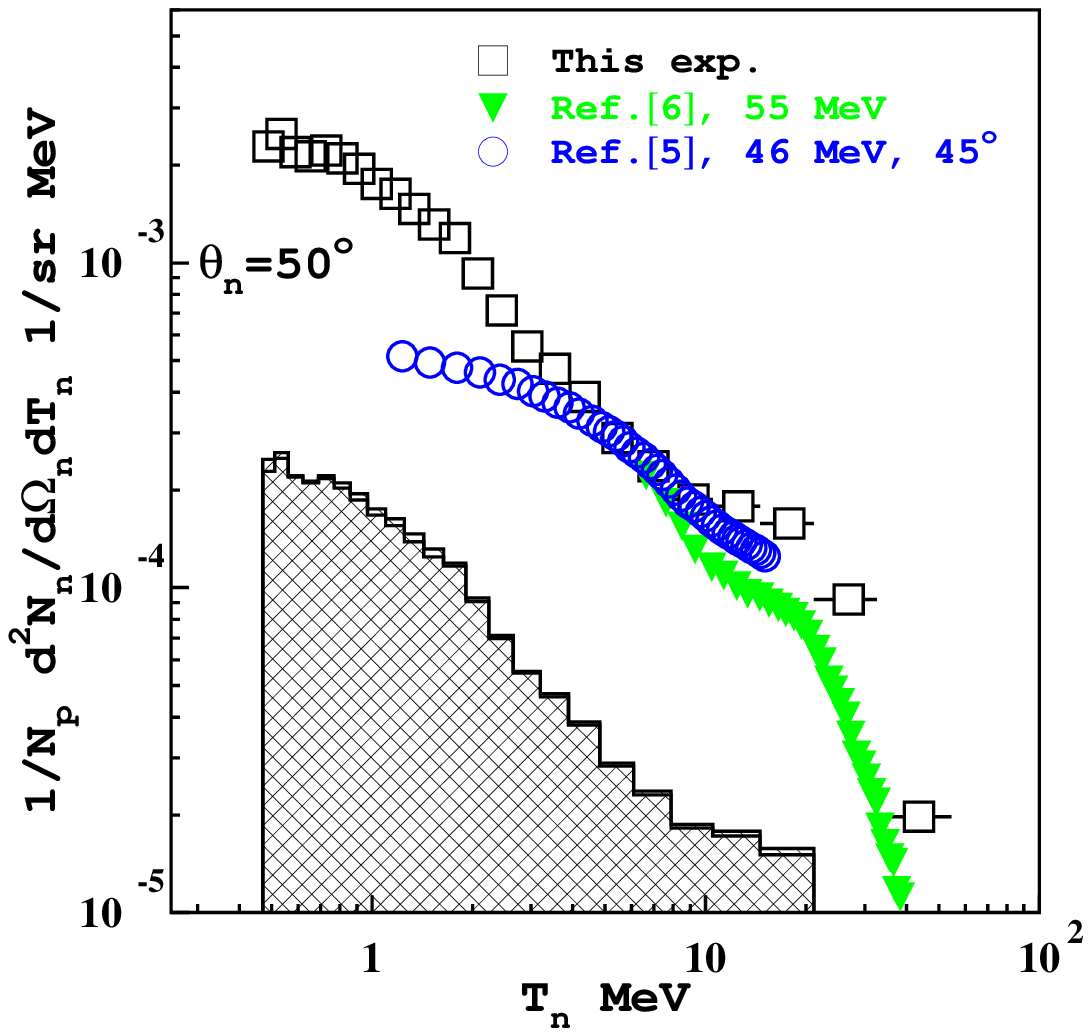}
\includegraphics[bb=2cm 5.5cm 20cm 23cm, scale=0.32]{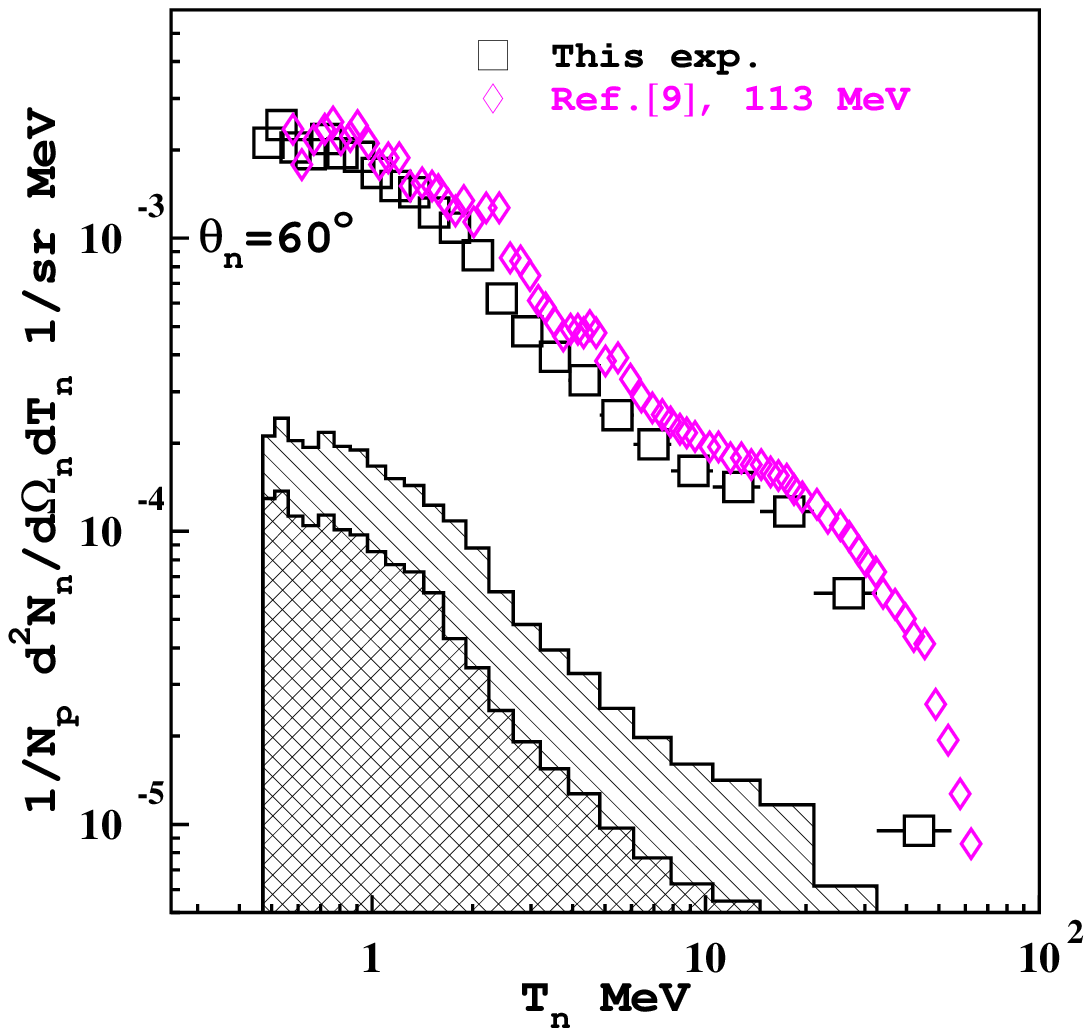}~%
\includegraphics[bb=2cm 5.5cm 20cm 23cm, scale=0.32]{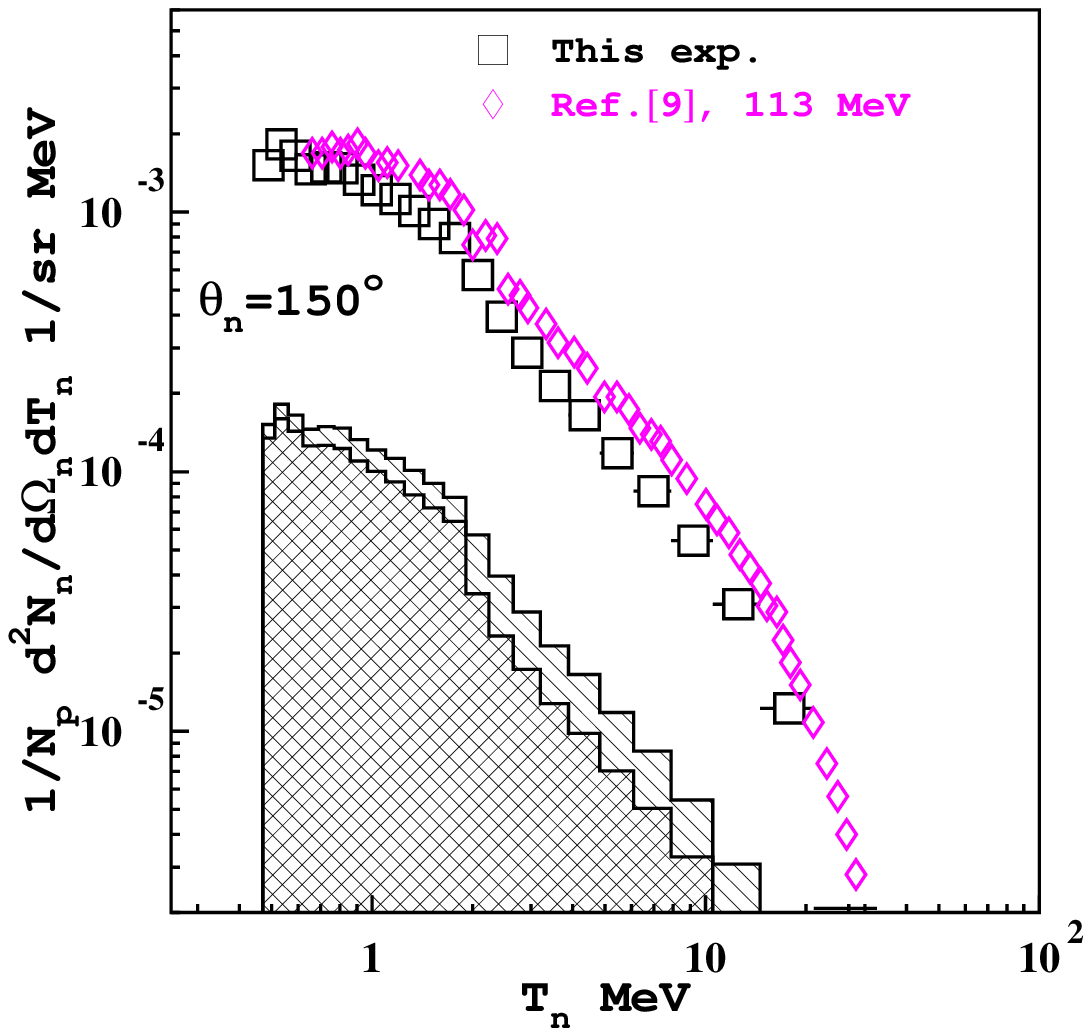}
\caption{\label{fig:cmp_world}(color online) Comparison of the measured yield to the world data from
Refs.~\cite{Almos77,Johnsen76,Waterman79,Meier88}.
The hatched histograms show systematic uncertainties due to
efficiency (right hatched) and normalization (left hatched).}
\end{center}
\end{figure*}

The comparison of the measured yields to world data at beam energies close to the
62 MeV of this experiment is shown in Fig.~\ref{fig:cmp_world}.
The data from Ref.~\cite{Almos77}, taken at almost the same beam energy, exhibit the same
high energy structure in the measured yield at small angles.
The data from Refs.~\cite{Johnsen76,Waterman79} taken at lower beam energies,
while exhibiting similar overall behavior, are suppressed at high $T_n$ due to the shrinkage
of the available phase space. Moreover, the data from Ref.~\cite{Waterman79},
which extend to lower neutron energies, exhibit a much milder ascent towards small $T_n$,
incompatible with our data.
Instead the data from Ref.~\cite{Meier88} taken at 113 MeV
have similar behavior to our points, but predictably lie slightly above.

\begin{figure*}[!ht]
\begin{center}
\includegraphics[bb=2cm 6.5cm 20cm 23cm, scale=0.32]{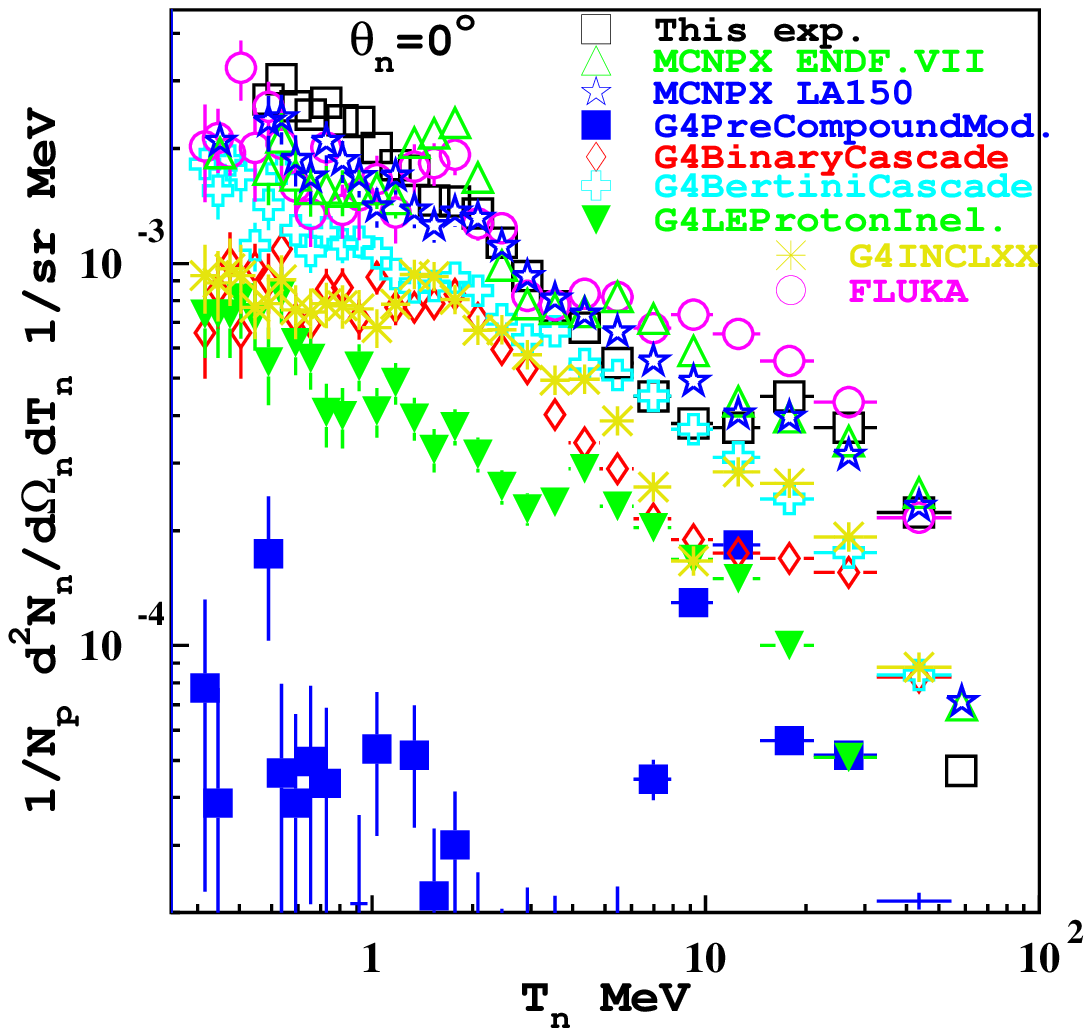}~%
\includegraphics[bb=2cm 6.5cm 20cm 23cm, scale=0.32]{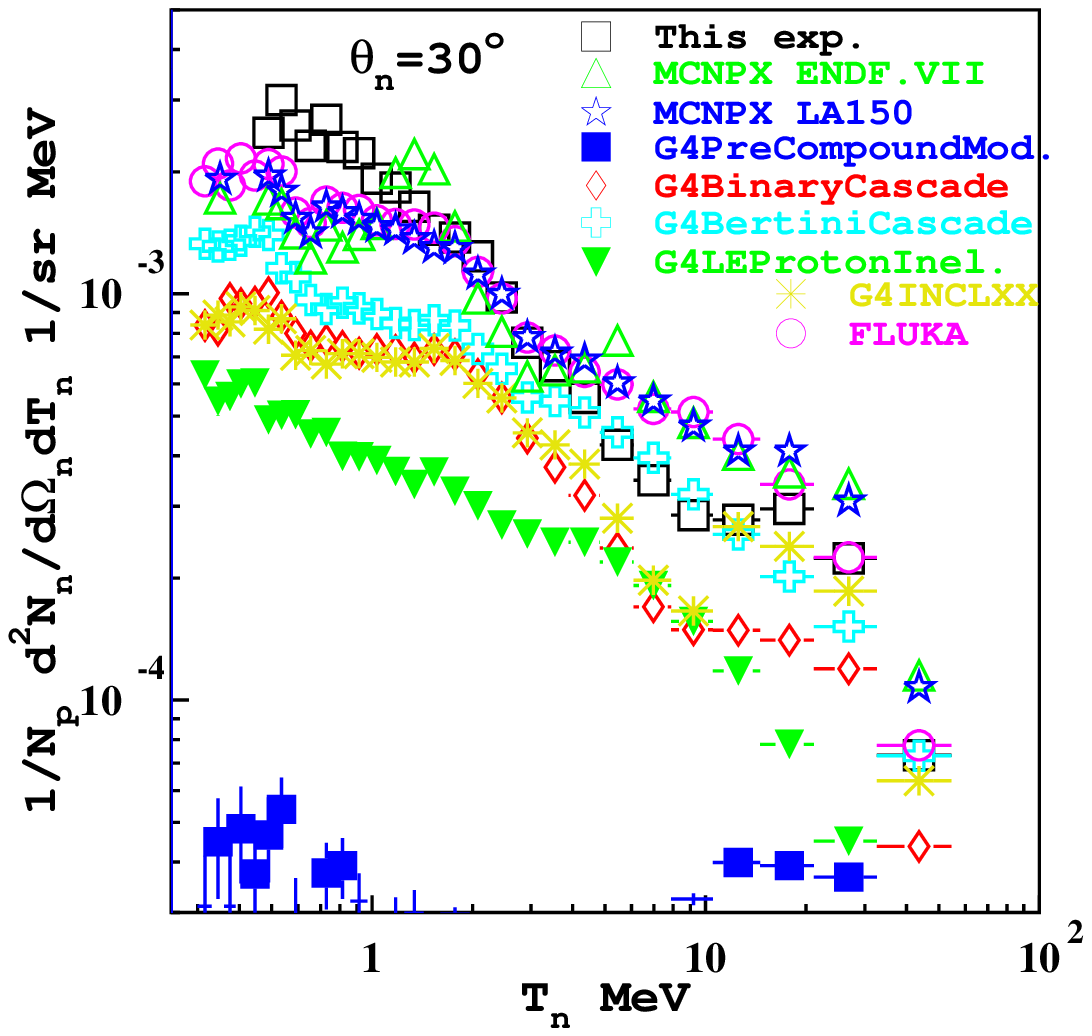}
\includegraphics[bb=2cm 6.5cm 20cm 24cm, scale=0.32]{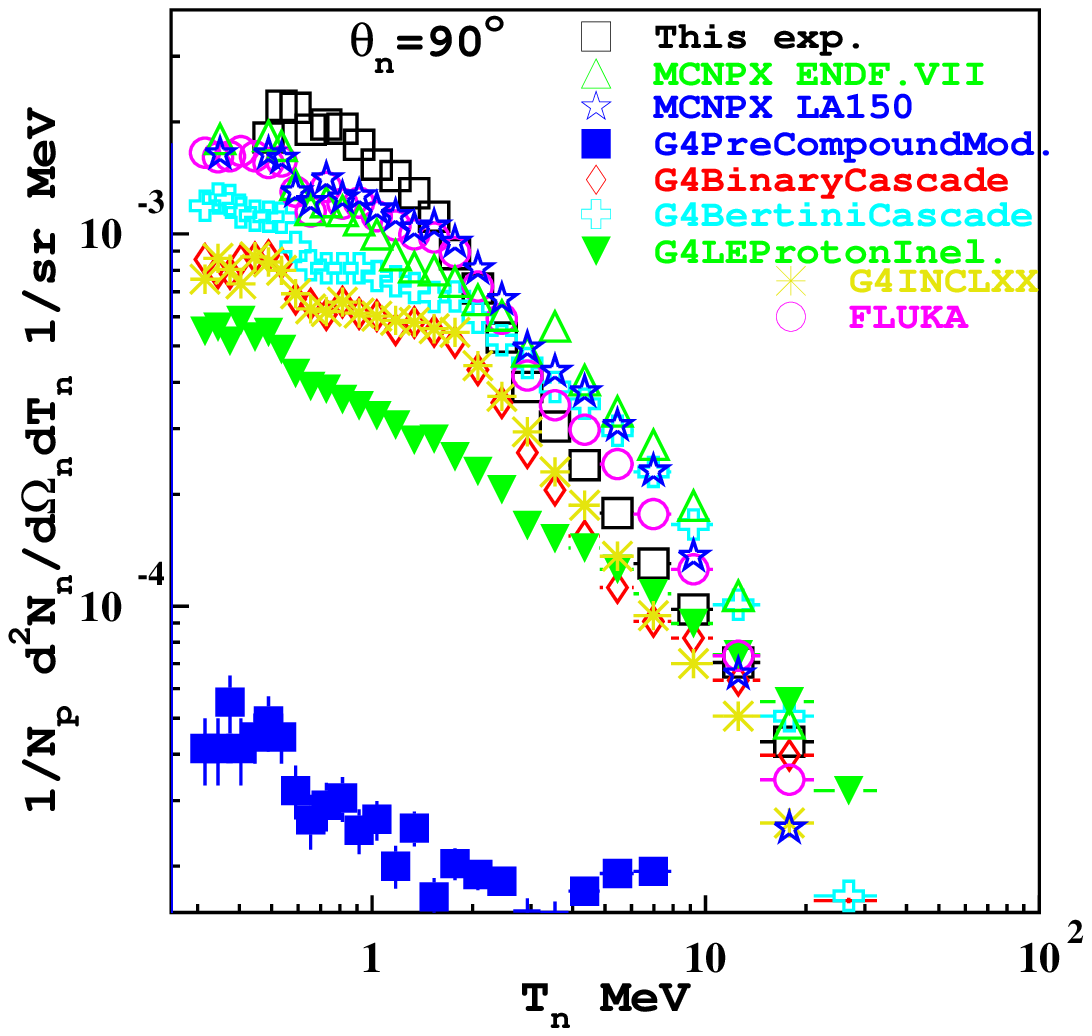}~%
\includegraphics[bb=2cm 6.5cm 20cm 24cm, scale=0.32]{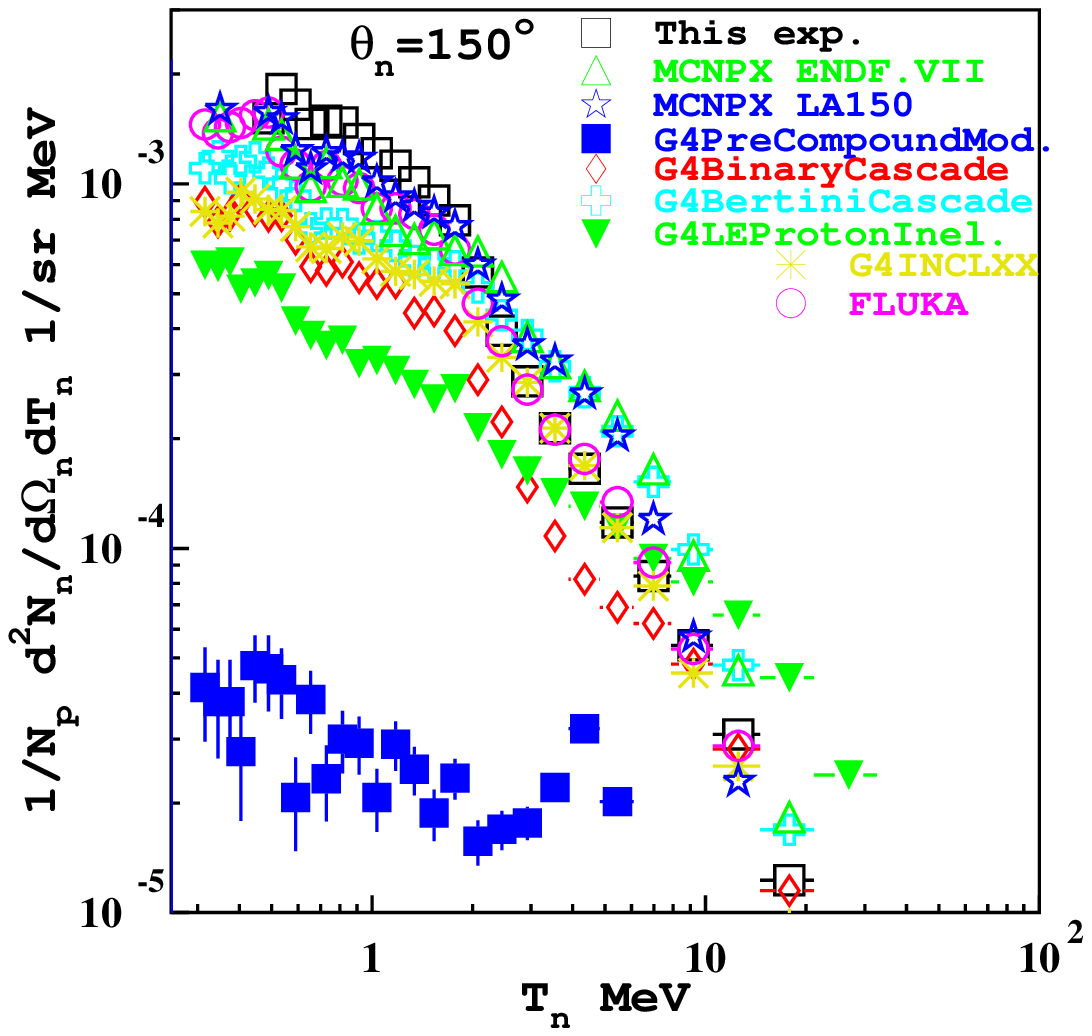}
\caption{\label{fig:cmp_sim}(color online) Comparison of the measured yields to the MCNP, FLUKA and Geant4 simulations at different angles.}
\end{center}
\end{figure*}

Our data were also compared to MCNP~\cite{mcnp5}, FLUKA~\cite{fluka1,fluka2} and Geant4~\cite{geant4} simulations performed using different physics libraries. The results are shown in Fig.~\ref{fig:cmp_sim}.
MCNP simulations have a different descending trend as compared to our data such that they underestimate significantly
(by a factor of 2 at small angles) the low energy part, while overestimating the high energy behavior
at large angles. Very similar distributions were obtained with FLUKA, which at small angles
had an overall higher magnitude.
Geant4 provides a broad choice of physics models to test against our data.
However, considering the energy range of the present experiment the set of applicable models
is reduced to five libraries: PreCompound model, Binary Cascade, Bertini Cascade, LEProtonInelastic and INCLXX.
The PreCompound model, mostly used to describe final state emission, was proposed also
as stand-alone model. Although it manages to form a sufficient number of $^{10}$B compound nuclei in the $p+^9Be$ interaction, the excess energy, instead of going into nucleus excitation, gets lost via emission of few high energy $\gamma$s. As a result the simulated yield is two orders of magnitude lower than the data.
LEProtonInelastic does not generate two major channels: $^8Be$ and 2 $\alpha$. It also frequently releases the excess energy through a high energy $\gamma$.
Binary Cascade has too high threshold (of about 30 MeV) for $^8Be$ production.
INCLXX model does not have this problem, but shows similar results.
The best Geant4 model for our kinematics, Bertini Cascade, gives results similar to MCNP and FLUKA, and also underestimates significantly the yield at low neutron energy.

%

\subsection{Systematic uncertainties}\label{sec:sys_err}
There are two major contributions to the systematic uncertainties in this experiment.
These were the overall normalization of the data and the knowledge of the liquid scintillator efficiency. We assumed that all other uncertainties were either negligible or folded into these two contributions.

The normalization was based on the charge accumulated on the Beryllium target and read out by a digital integrator. The precision of the integrator was better than 2 pA, but in some runs corrections to the measured charge were necessary as discussed in section~\ref{sec:fc_charge}. Therefore we evaluated the normalization systematic uncertainty as the maximum between 2 pA value and the correction applied.

The efficiency systematic uncertainties were evaluated from the comparison of Geant4 simulations to the dedicated measurement described in section~\ref{sec:effic}. From this comparison the standard deviation of the data from Geant4 was extracted in the range 0.5-3 MeV for the small detectors and 2-10 MeV for the large ones. The deviation resulted to be 10\% for both types of detectors. This value was taken as an estimate of efficiency systematic uncertainty.

\subsection{Low energy detectors}
At low neutron energy $T_n<0.47$ MeV liquid scintillators were not efficient enough to determine
the neutron flux. This energy range was measured by two different detectors:
one was a $^3$He tube (Canberra 150NH50/5A) and the other a silicon detector (Micron MSX09) covered with a $^6$Li converter.
Both preamplified detector pulses were amplified by an Ortec 672 spectroscopy amplifiers
and then digitized by the same DAQ system using a SILENA 4418 peak sensing ADC.
In few special runs these detectors were installed on the same supports used for liquid scintillators.
Assuming the low energy neutron flux to be isotropic we measured only few angles:
70 and 90 degrees for $^3$He tube and 130 degrees for $^6$Li-silicon detector.
The event rates were fairly low, hence the distance to the target was chosen to be 64 cm
for $^3$He tube and 69 cm for $^6$LiF-silicon detector.

The $^3$He tube energy response to thermalized neutrons from an AmBe source was first analyzed for the purpose of ADC calibration. Using this calibration, energy spectra of the data taken during the experiment were obtained.

\begin{figure}[!ht]
\begin{center}
\includegraphics[bb=2cm 6cm 20cm 23cm, scale=0.35]{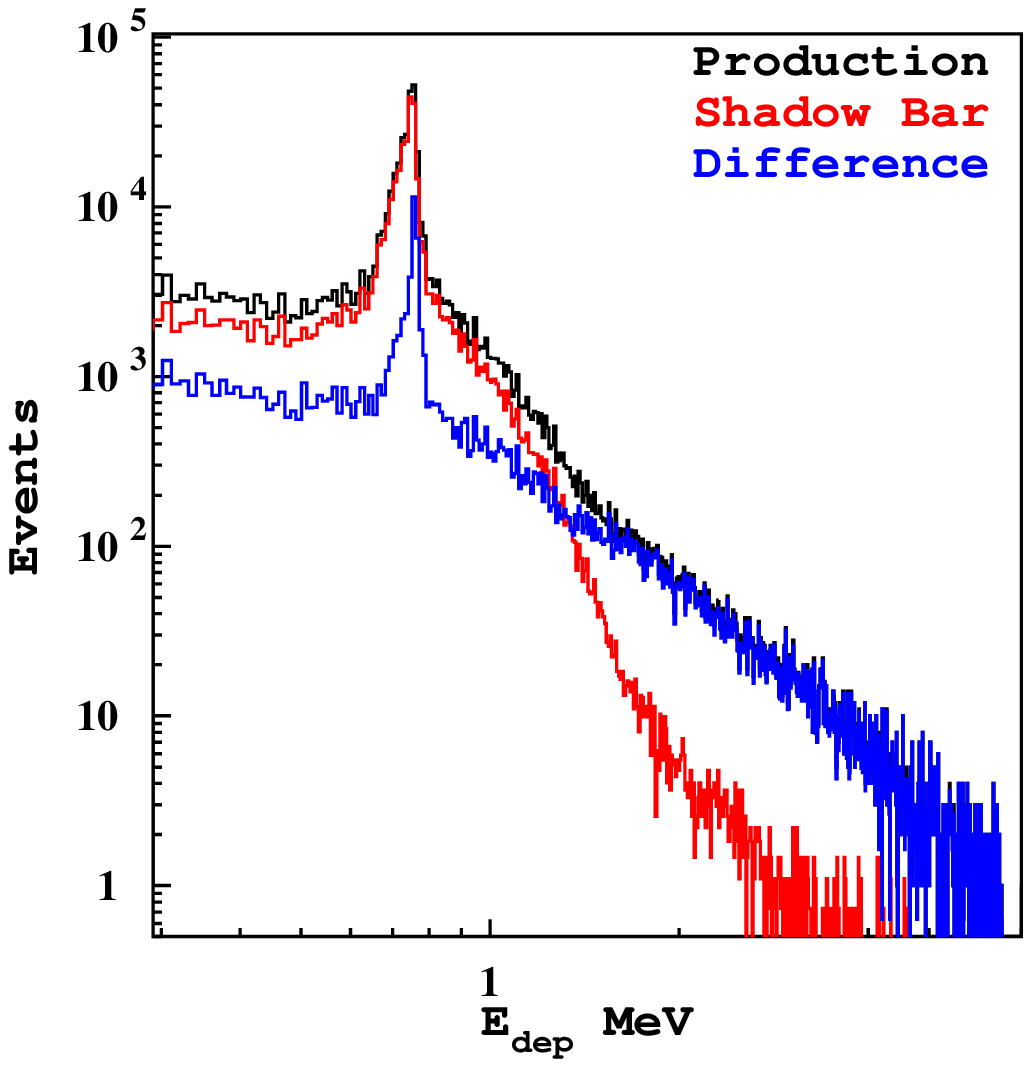}
\caption{\label{fig:he3_shadow_bar}(color online) Normalized $^3$He detector spectra measured at 90 degrees.
The black histogram shows the production run data, red histogram represents the spectrum obtained
in the shadow bar run and their difference is given by the blue histogram.}
\end{center}
\end{figure}

The contribution of the environmental background of neutrons scattered from the surrounding materials was suppressed by wrapping the lateral cylinder surface with a 1 mm thick Cd foil,
absorbing completely neutrons with energies below 0.5 eV.
In order to remove the remaining environmental background shadow bar measurements,
using a 1 mm thick Cd foil together with the 50 cm steel cylinder,
were subtracted from the data.
The environmental background subtraction is shown in Fig.~\ref{fig:he3_shadow_bar}. The background
contribution drops rapidly for deposited energies above 1.5 MeV, corresponding to neutrons with
energies higher than 0.8 MeV. Since the $^3$He tube geometry
was not optimized for this particular experiment the overall background contribution reached 70\%,
leading to large systematic uncertainties.

\begin{figure}[!ht]
\begin{center}
\includegraphics[bb=2cm 6cm 20cm 23cm, scale=0.35]{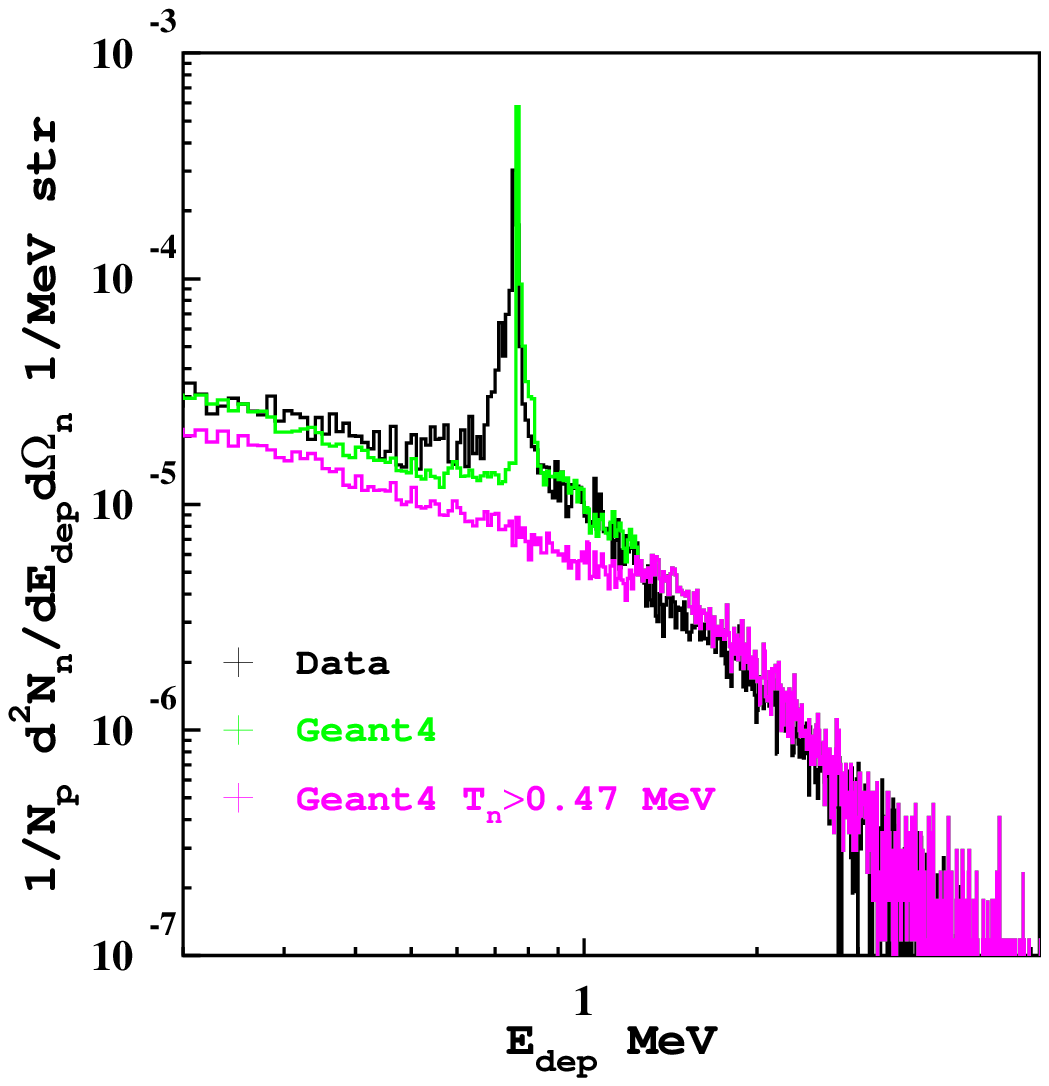}
\caption{\label{fig:he3_edep}(color online) Subtraction of the fast neutron contribution,
as measured by liquid scintillators ($T_n>0.47$ MeV), from the measured $^3$He spectra
using Geant4 Monte Carlo simulations.
The black histogram shows the $^3$He data, green histogram represents the Geant4 simulations of entire spectrum
and magenta histogram gives the contribution of fast neutrons measured by liquid scintillators.
}
\end{center}
\end{figure}

The obtained spectrum was divided in five bins from 0 to 0.47 MeV according to the energy resolution estimated from Geant4 simulations. Then an unfolding procedure was applied to the data. First we simulated in Geant4 the $^3$He tube response to the neutron flux measured with liquid scintillators and subtracted it from the data
as shown in Fig.~\ref{fig:he3_edep}.

\begin{figure}[!ht]
\begin{center}
\includegraphics[bb=2cm 6cm 20cm 23cm, scale=0.35]{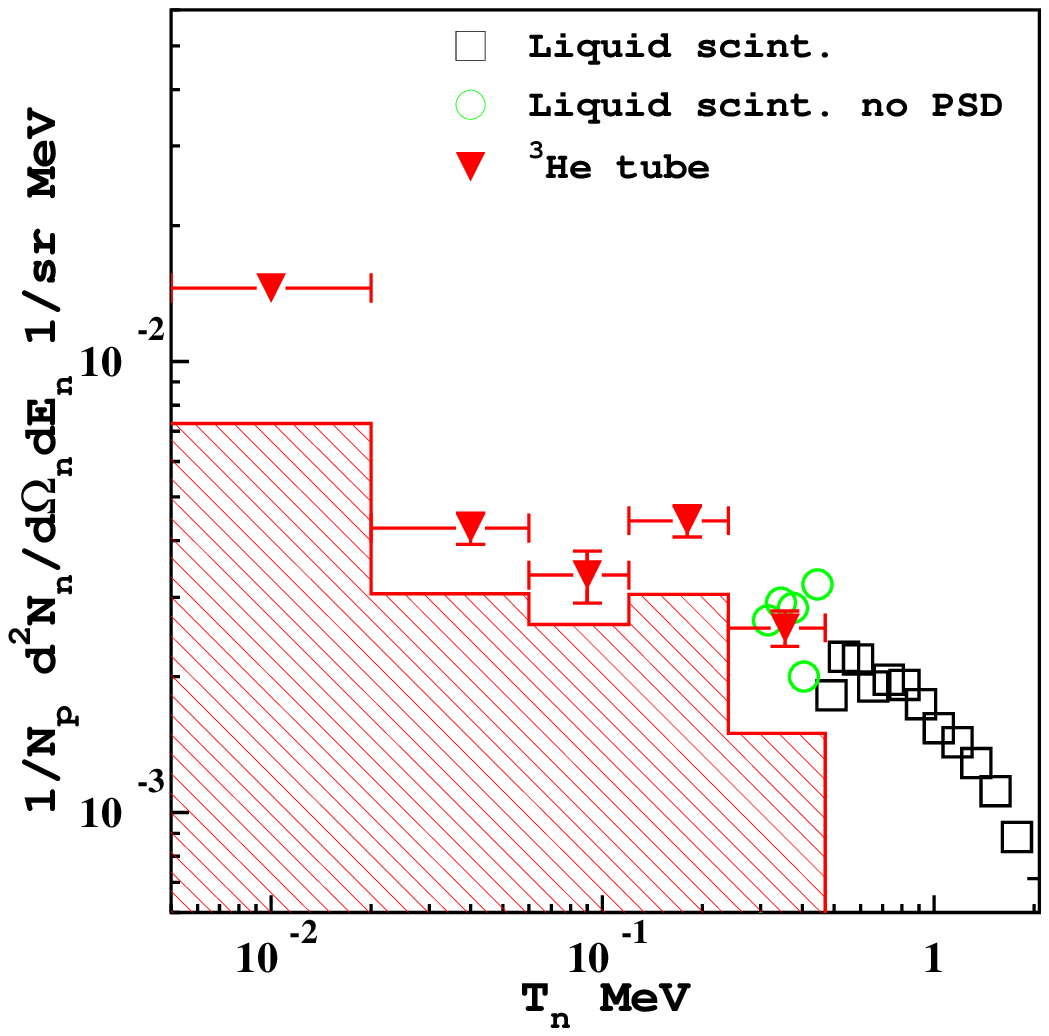}
\caption{\label{fig:le_extrapol}(color online) Energy dependence of the neutron yield at low energy
measured by the $^3$He tube in comparison with liquid scintillator data.
Red triangles represent $^3$He data, while red dashed area gives its systematic uncertainty.
Black squares and green circles show liquid scintillator data with and without PSD.}
\end{center}
\end{figure}

Then we extrapolated this flux to lower energies via $T_n^\beta$ dependence, with $\beta$ being an adjustable parameter, and extracted neutron yield in defined kinematic bins using Geant4 simulations of efficiency. This step was repeated few times adapting at every iteration the simulated neutron yield to the extracted one. Once the extraction procedure converged, we obtained the final neutron yield at low energies shown in Fig.~\ref{fig:le_extrapol} in comparison to liquid scintillator data.

The highest point of $^3$He data was found to be in good agreement with liquid scintillator measurements.
In particular, the data from liquid scintillator without PSD extend down to 0.3 MeV with somewhat larger systematic uncertainties.
Although the estimated systematic uncertainties of $^3$He data resulted to be $>50$\%, the obtained data allowed to constrain the missing part of the neutron energy spectra.
The systematic uncertainty of $^3$He data contributes for 50\% to the uncertainty of the integrated neutron yield, although in absolute value it corresponds to only 3\% of the yield.

The $^6$LiF-silicon detector data confirmed the neutron yield extracted from liquid scintillator measurements
and put compatible upper limits on $^3$He data. However, due to large silicon active layer thickness it was mainly sensitive to high energy neutrons.

\subsection{Integrated yield}
\begin{figure}[!ht]
\begin{center}
\includegraphics[bb=2cm 6cm 20cm 23cm, scale=0.35]{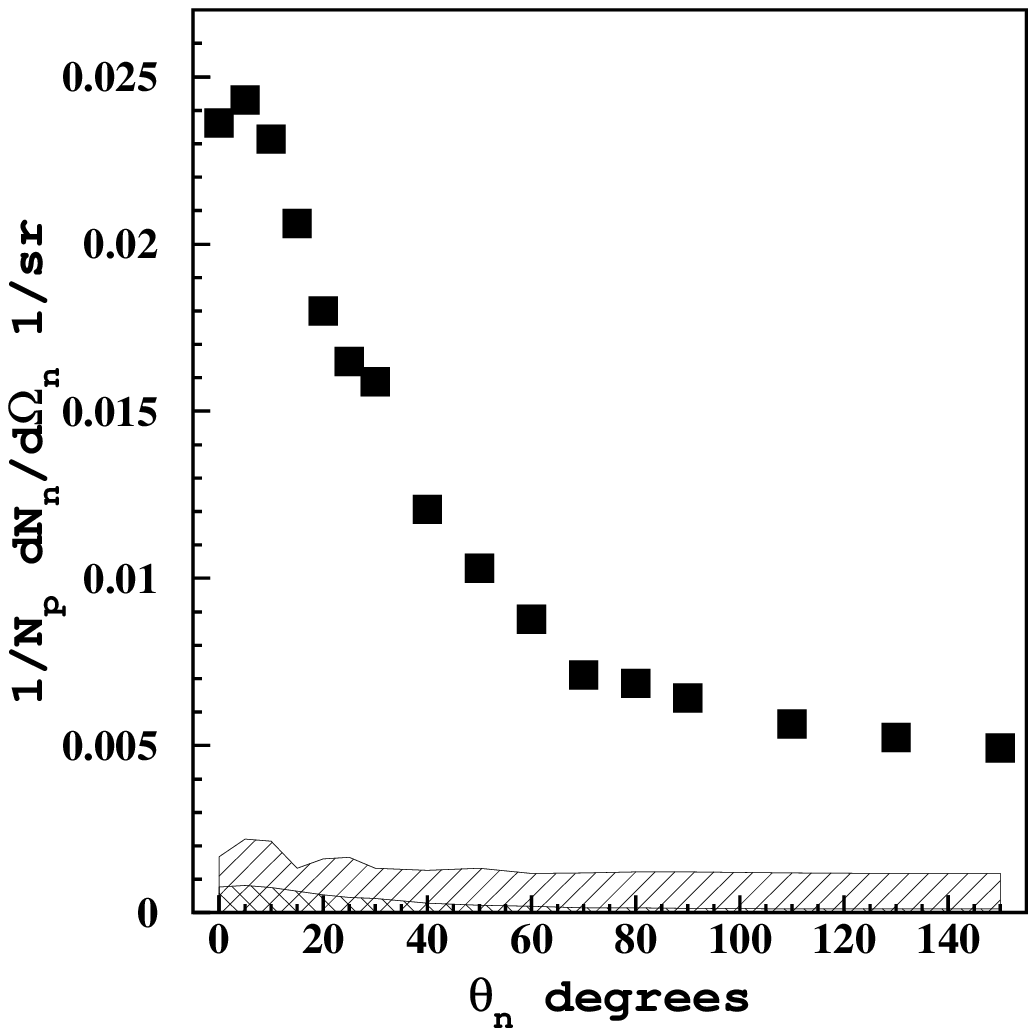}
\caption{\label{fig:int_yield} Angular dependence of the integrated neutron yield.
The hatched histograms show systematic uncertainties due to
efficiency (right hatched) and normalization (left hatched).}
\end{center}
\end{figure}

Combining liquid scintillator data and $^3$He tube data the complete neutron energy coverage was achieved. Thus the integrated neutron yield was calculated separately for each angular setting.
This was performed assuming that the low-energy part of the spectrum was isotropic, supported by the low energy points from liquid scintillator detectors.
The obtained angular distribution of the integrated neutron yield is shown in Fig.~\ref{fig:int_yield}.

The integrated neutron yield falls rapidly with the angle up to about 90 degrees, while at larger angles
the distribution is almost flat. This suggests a large contribution of the process where the incident
proton kicks off the loosely bound neutron via a quasi-elastic scattering.
Instead, at large angles the dominant process is, perhaps, of evaporative nature.
By integrating, in turn, the neutron yield in Fig.~\ref{fig:int_yield} over the angle, the total neutron
yield $N_n/N_p=0.0987\pm0.0003_{stat.}\pm0.0053_{sys.}$ n/p was obtained. This value can be compared to the MCNP simulations $N_n/N_p=0.103$ n/p (ENDF VII) or $N_n/N_p=0.096$ n/p (LA150) performed for the same beam energy, and to the experimental data from Ref.~\cite{Tilquin05} $N_n/N_p=0.110\pm0.007$ n/p at 65 MeV. The total yield from Ref.~\cite{Tilquin05} was measured by thermalization of produced
neutrons and their subsequent capture.

\section{Conclusions}\label{sec:conclusions}
Fully differential neutron yields produced by 62 MeV proton beam impinging on a thick, fully absorbing $^9$Be target were measured in almost complete energy and angular ranges.
The precision of the data is limited by systematic uncertainties on the detector efficiency and by the absolute normalization. These uncertainties of the yields are of the order of 10\%. A better precision on detector efficiency could be achieved by means of a precisely calibrated neutron source, while the normalization uncertainty could be improved by using a higher beam intensity.

The obtained data are in agreement with previous experiments in 60-70 MeV beam energy range, extending considerably the kinematic coverage. The overall yield, integrated over neutron energy and angle, is in good agreement with the measurement based on thermalization and capture of produced neutrons~\cite{Tilquin05}.
The comparison of our data to MCNP and FLUKA simulations showed some deviations in angular
and neutron energy behaviors, although the the integrated yield values were found to be consistent.
The comparison to various Geant4 models exhibited their limitations in the low beam energy domain.

The presented data will be used in the design of the ADS core~\cite{infn_e_ads} to describe
its neutron source.
These data can also be useful for the development of the neutrino source proposed recently in Ref.~\cite{isodar}.


\section*{Acknowledgement}
Authors would like to acknowledge the excellent support provided during the experiment
by the accelerator staff and technical services of Laboratori Nazionali del Sud.
Authors are also grateful to Andrea Bersani for performing ANSIS simulations of the carbon fiber beamline
and to Barbara Caiffi for carrying out MCNP simulations.
This work was supported by the Istituto Nazionale di Fisica Nucleare INFN-E project.


\bibliographystyle{elsarticle-num}
\bibliography{lns_exp2012}

\end{document}